\documentclass[fleqn,usenatbib]{mnras}

\usepackage{mathptmx}
\usepackage{txfonts}

\usepackage[T1]{fontenc}
\usepackage{ae,aecompl}

\usepackage[dvips]{color}

\usepackage{graphicx}
\usepackage{lscape}

\def\ergscm{erg~s$^{-1}$~cm$^{-2}$}

\def\lum{erg s$^{-1}$}

\def\taut{\tau_{\rm T}} 
\def\sigmat{\sigma_{\rm T}} 
\def\sgra{Sgr~A$^{\star}$}

\def\apj{ApJ}

\def\aap{A\&A}
\def\mnras{MNRAS}
\def\apjs{ApJS}

\def\apjl{ApJLett}
\def\procspie{\ref@jnl{Proc.~SPIE}}   

\def\arches{Arches cluster}
\def\x1{NGC~5643~X-1}
\def\apec{APEC}
\def\nu{\textit{NuSTAR}}
\def\xmm{\textit{XMM--Newton}}

\title[\nu\ and \xmm\ observations of the Arches cluster in 2015]{\nu\ and \xmm\ observations of the Arches cluster in 2015: fading hard X-ray emission from the molecular cloud} 
\author[Krivonos et al.]{Roman\,Krivonos$^{1}$\thanks{E-mail:
    krivonos@iki.rssi.ru}, Ma{\"i}ca Clavel$^2$, JaeSub Hong$^{3}$,
  Kaya Mori$^{4}$, Gabriele Ponti$^{5}$ \and
  Juri Poutanen$^{6,7}$, Farid Rahoui$^{8,9}$, John Tomsick$^{2}$ and
  Sergey Tsygankov$^{6}$ \\
$^{1}$Space Research Institute of the Russian Academy of Sciences,
Profsoyuznaya Str. 84/32, 117997 Moscow, Russia;\\
$^{2}$Space Sciences Laboratory, 7 Gauss Way, University of
California, Berkeley, CA 94720-7450, USA;\\
$^{3}$Harvard-Smithsonian Center for Astrophysics, 60 Garden Street,
Cambridge, MA 02138, USA; \\
$^{4}$Columbia Astrophysics Laboratory, Columbia University, New York, NY
10027, USA; \\
$^{5}$Max-Planck-Institut f\"ur extraterrestrische Physik, Giessenbachstrasse 1, Garching, 85748, Germany;\\
$^{6}$Tuorla Observatory, Department of Physics and Astronomy, University of Turku, V\"ais\"al\"antie 20, FI-21500 Piikki\"o, Finland;\\
$^{7}$Nordita, KTH Royal Institute of Technology and Stockholm University, Roslagstullsbacken 23, SE-10691 Stockholm, Sweden;\\
$^{8}$European Southern Observatory, Karl-Schwarzschild-Str. 2,
D-85748 Garching bei M\"unchen, Germany; \\
$^{9}$Harvard University, Department of Astronomy, 60 Garden street,
Cambridge, MA 02138, USA; \\
}

\begin{document}
\label{firstpage}
\pagerange{\pageref{firstpage}--\pageref{lastpage}}
\maketitle

\begin{abstract}

  We present results of long \nu\ (200~ks) and \xmm\ (100~ks)
  observations of the Arches stellar cluster, a source of bright
  thermal ($kT\sim$2~keV) X-rays with prominent Fe~XXV~K$\alpha$~6.7 keV
  line emission and a nearby molecular cloud, characterized by an
  extended non-thermal hard X-ray continuum and fluorescent Fe
  K$\alpha$~6.4 keV line of a neutral or low ionization state material
  around the cluster.  Our analysis demonstrates that the non-thermal
  emission of the Arches cloud underwent a dramatic change, with its
  homogeneous morphology, traced by fluorescent Fe K$\alpha$ line
  emission, vanishing after 2012, revealing three bright clumps.  The
  declining trend of the cloud emission, if linearly fitted, is
  consistent with half-life decay time of $\sim8$~years. Such strong
  variations have been observed in several other molecular clouds in
  the Galactic Centre, including the giant molecular cloud Sgr~B2,
  and point toward a similar propagation of illuminating fronts,
  presumably induced by the past flaring activity of \sgra.
 
\end{abstract}


\section{Introduction}

The Arches cluster is a massive star cluster with a core that is about
$9''$ ($\sim0.35$~pc at 8~kpc) in radius \citep{figer1999}, containing
more than 160 O-type stars \citep{figer2002}. It is
located in the inner Galactic Center (GC) region at the projected
angular distance of $11'$ from \sgra. \textit{Chandra} observations
of the Arches cluster region revealed a complicated picture of the
X-ray emission, showing the presence of spatially separated thermal and
non-thermal emission components. The thermal emission is thought to
originate in the cluster's core from multiple collisions between strong
winds of massive stars \citep{zadeh2002,wang2006,capelli11a}.  Diffuse
non-thermal X-ray emission has been detected from a broad region
(``clouds'') around the cluster
\citep{wang2006,tsujimoto2007,capelli11b,T12,K14}. The non-thermal
nature of this extended radiation is revealed by its strong
fluorescent Fe~K$\alpha$~$6.4$~keV line emission, presumably coming
from material that is neutral or in a low ionization state.

Two physical mechanisms to produce fluorescent emission in the
molecular clouds are generally discussed. The first implies reflection
of incoming hard X-rays where K-shell photoionization and the
subsequent fluorescence produce a strong Fe K$\alpha$ line with an
equivalent width $EW\gtrsim 1$~keV
\citep{sunyaev1993,markevitch1993,koyama1996,sunyaev1998}. Additionally,
the Compton scattering of high-energy photons results in a reflection
hump around $20-30$~keV. The source of the incoming X-rays might be
associated with a nearby X-ray source
\citep[e.g.][1E1740.7-2942]{churazov1993_1e1740} or with past activity of
\sgra, as suggested by \cite{sunyaev1993} to explain the hard
X-ray emission of the giant molecular cloud Sgr~B2 in the GC region,
leading to prediction of the bright 6.4~keV line later confirmed by
the observations
\citep{koyama1996,sunyaev1998,murakami2000,revnivtsev2004a,terrier2010,shuo2015}.
A long outburst from \sgra\ lasting more than 10 years and ending a
few hundred years ago could explain the Sgr~B2 emission
\citep{koyama1996,terrier2010}.  The hypothesis of past \sgra\ flaring
activity is supported by the discovery of a propagation of Fe K echos
in the Central Molecular Zone (CMZ), presumably from a source far away
from the clouds
\citep{ponti2010,ponti2013,clavel2013,ryu2013}. Alternative
explanation suggests the propagation of cosmic-ray (CR) particles
within the molecular clouds, which can produce hard X-rays and Fe
K$\alpha$ emission \citep{valinia2000,dogiel2009,T12,dogiel2014},
however CR scenarios fail to reproduce the fast variability that is
observed in numerous clouds of the CMZ \citep{terrier2017}.

Based on the archival \textit{XMM} data of the GC region,
\cite{capelli11b} and \cite{T12} constructed Fe~K$\alpha$ emission
line maps of the Arches cluster region at $6.4$~keV and showed that
this emission extends well beyond the cluster's
core. \cite{capelli11b} argue that the observed Fe~K$\alpha$ line flux
and the high value of the equivalent width
(\textit{EW}$_{\textrm{\tiny 6.4 keV}}$) suggest an origin of the
fluorescence in the photoionization of the cloud by X-ray photons,
although excitation by CR is not specifically excluded. \cite{T12}
(hereafter T12) suggested that the Fe~K$\alpha$~$6.4$~keV line emission
observed around the Arches cluster is likely produced by bombardment
of molecular gas by low-energy cosmic-ray protons (LECR{\it
  p}). The authors suggest that the required large flux of low-energy CR
particles could be produced in the ongoing supersonic collision
between the star cluster and a nearby molecular cloud.

Using the first hard X-ray focused observation of the Arches cluster
region, performed with \nu, we showed in our previous work
\citep[][hereafter K14]{K14} that the continuum emission in the
$10-20$~keV band is significantly detected around the Arches cluster
with a spatial morphology consistent with the Fe~K$\alpha$ fluorescent
line emission.

\citet[][hereafter C14]{clavel2014} analysed the long-term behavior of
the Arches cluster cloud (or simply the Arches cloud) over 13 years
and reported a 30\% decrease ($4\sigma$ confidence) in Fe K$\alpha$
line and continuum flux of the cloud emission, providing significant
evidence for the X-ray reflection scenario. Despite this success, the
question of determining the illuminating hard X-ray source and
reflection geometry remains open. The source cannot be within the
Arches cluster, according to constraints drawn from the Fe K$\alpha$
line flux (T12) and hard X-ray continuum (K14), leaving the
possibility for putative past activity of \sgra. In this work we track
further the evolution of the flux of the Fe K$\alpha$ line with \xmm\
observations and measure the shape of the hard X-ray continuum with
\nu.

Although the CR-only scenario is practically ruled out by the observed
variability for most of the measured non-thermal emission, the origin
of the remaining faint emission is not so obvious. As the non-thermal
flux decreases, we might now be observing a significant contribution
from a putative steady background level, which could be created by a
different process than the one responsible for the bulk of variable
emission. In this paper, we repeat the spectral fit performed by K14
to derive updated constraints for both the X-ray photoionization and
the LECR emission models based on the spectral shape above 10 keV
using the \nu\ data taken in 2015.

The paper is structured as follows.  In Sect.~\ref{sec:data} we
describe observations of the Arches cluster with \xmm\ and \nu\ in
2015 and outline the corresponding data analysis. Spectral modelling
of the Arches cluster core and cloud emission is presented in
Sect.~\ref{sec:spec}. Discussion of the results obtained and the
summary can be found in Sect.~\ref{sec:discussion} and
\ref{sec:summary}, respectively.  Additional materials (e.g.,
serendipitous detection of the Sgr A complex and a description of
2D image spectral extraction procedure) are placed in the Appendix.

\section{Observations and Data Analysis}
\label{sec:data}

\begin{table}
\noindent
\centering
\caption{List of the X-ray observations of the \arches\ with \xmm\ and
  \nu\ used in this work}\label{tab:log} 
\centering
\vspace{1mm}
  \begin{tabular}{|c|c|r|c|c|c|c|}
\hline\hline
Mission & Date & ObsID & Exp. & Frac.$^{*}$ \\
& & & (ks) & (\%) \\
\hline
\xmm & 2015-09-27 & 0762250301 &   112.0 & 78/84/60\\
\nu & 2015-10-19 & 40101001002 & 107.2 & 100/0 \\
\nu & 2015-10-25 & 40101001004 & 107.8  & 100/0 \\
\hline
\end{tabular}\\
\begin{flushleft}
$^{*}$Good time fraction for \xmm\ MOS1/MOS2/PN and \nu\ FPMA/FPMB,
respectively; Note that \nu\ FMPB data were completely rejected due to
stray-light contamination (Sect.~\ref{sec:data:nustar}). 
\end{flushleft}
\vspace{3mm}
\end{table}

\subsection{\xmm}
\label{sec:data:xmm}

The \xmm\ data reduction was carried out using the \textit{XMM-Newton}
Extended Source Analysis Software \citep[ESAS,][]{snowden2008}
included in version 14 of the \textit{XMM-Newton} Science Analysis
Software (SAS).  We followed the procedure described by C14, using the
SAS \textit{emchain} and \textit{epchain} scripts to produce the
calibrated event lists and ESAS mos-filter and pn-filter to exclude
periods affected by soft proton flaring.

The \xmm\ source and astrophysical background spectra were extracted
from these clean event lists with the ESAS \textit{mos-spectra} and
\textit{pn-spectra} tasks. For the `Cloud' region, we also extracted
the corresponding quiescent particle background (QPB) from the filter
wheel closed event lists provided by the ESAS calibration database and
normalized them to the level of QPB in the observations, using
\textit{pn\_back} and \textit{mos\_back}. Spectrum counts were grouped
to have at least thirty counts per bin so that we can validly use
chi-squared statistics for fitting (Sect.~\ref{sec:spec}).

\begin{figure*}
\includegraphics[width=0.42\textwidth]{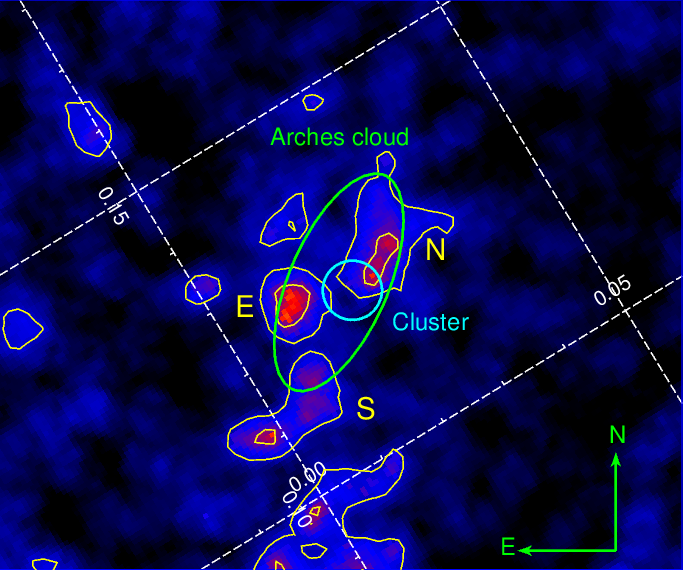}
\includegraphics[width=0.42\textwidth]{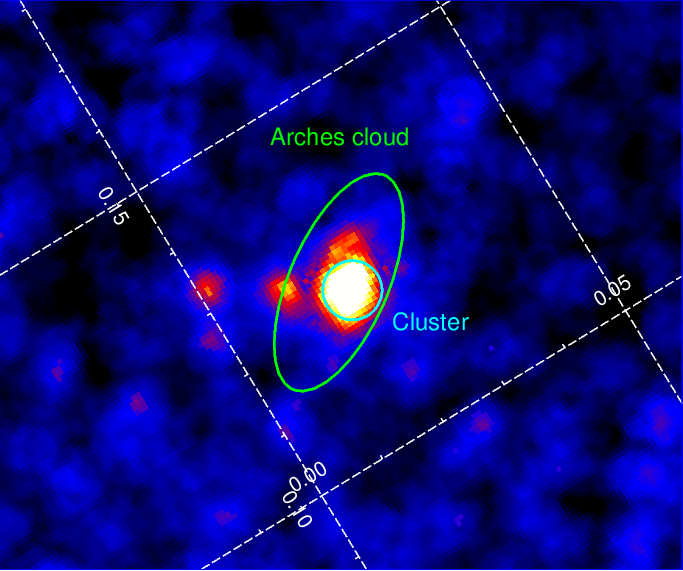}
\caption{\xmm\ K$\alpha$ line mosaic image of the Arches cluster
  region of the iron at 6.4~keV (left) and 6.7~keV (right) obtained in
  $6.32-6.48$ and $6.62-6.78$~keV energy bands, respectively. The
  images are continuum subtracted and adaptively smoothed. The
  contours in the left image are shown to highlight bright clumps of
  6.4~keV emission. The compass sign shows the
  images alignment in equatorial coordinates, with North up and East
  to the left.}\label{fig:xmm:arches}
\end{figure*}

To produce the continuum-subtracted images we followed the optimized
procedure presented by \citet{terrier2017}, which can be summarized as
follow. For each instrument we produced the count, background and
exposure maps with initial resolution of 2.5\,arcsec, using
\textit{mos-spectra}, \textit{pn-spectra}, \textit{mos\_back} and
\textit{pn\_back} in three energy bands: 4.7--6.3, 6.32--6.48 and
6.62--6.78\,keV. To estimate the continuum underlying the emission
lines we assumed the following spectral model: a power-law of photon
index $\Gamma=2$ plus a thermal plasma of temperature $kT=6.5$\,keV
(with abundance equal to solar), both components being absorbed by a
column density $N_{\rm H} = 7 \times 10^{22}$\,cm$^{-2}$. For each
pixel, the normalizations of these two components were derived using
the continuum emission within 4.7--6.3\,keV and the line+continuum
emission at 6.7\,keV, allowing to subtract both the continuum emission
underlying the 6.4\,keV line and the non-thermal emission underlying
the 6.7\,keV line.  After taking into account the different
efficiencies of the three EPIC detectors, the individual instrument
maps were then combined to create the corresponding
continuum-subtracted flux maps. Fig.~\ref{fig:xmm:arches} shows Fe
K$\alpha$ images at 6.4~keV and 6.7~keV obtained, respectively, in
$6.32-6.48$~keV and $6.62-6.78$~keV energy bands. As expected, the
cluster's core is bright at 6.7~keV line revealing strong thermal
radiation in the stellar cluster.

\begin{table}
\begin{center}
  \caption{Definitions of the sky regions.\label{tab:regions}}
\begin{tabular}{cccccc}
\hline
 RA (J2000) & Dec (J2000) & Parameters\\
\hline
\multicolumn{3}{c}{Cluster (circle)} \\
$17^h45^m50.3^s$ & $-28^{\circ}49'19''$ & $15''$\\
\hline
\multicolumn{3}{c}{Cloud (ellipse)} \\
$17^h45^m51.0^s$ & $-28^{\circ}49'16''$ & $25'',
59'', 155^{\circ}$ \\
$17^h45^m50.3^s$ & $-28^{\circ}49'19''$ & $15''$ (excl.)\\
\hline
\multicolumn{3}{c}{Arches cluster complex (circle)} \\
$17^h45^m50.3^s$ & $-28^{\circ}49'19''$ & $50''$\\
\hline
\multicolumn{3}{c}{Background (annulus)} \\
$17^h45^m50.3^s$ & $-28^{\circ}49'19''$ & $130''$\\
$17^h45^m50.3^s$ & $-28^{\circ}49'19''$ & $70''$ (excl.)\\
\hline
\end{tabular}
\end{center}
Notes. Central position and radius for circular regions, and semi-minor/major axes and rotation angle for the
elliptical regions. The rotation is defined counter clockwise relative
to North (upward). Background region is shown in
Fig.~\ref{fig:nustar:arches}. ``Cluster'' and ``Cloud'' regions are illustrated
in Fig.~\ref{fig:xmm:arches}. The excluded regions are marked as ``excl.''.
\end{table}

The Arches cluster core thermal emission located within the circular
region of $R\sim15''$ is embedded in the elongated non-thermal
emission of the cloud with dimensions of $\sim25''\times59''$. In order to
perform spatial and spectral analysis consistent with recent studies
of the Arches cluster region (T12,~K14,~C14), we adopt the same sky
regions to describe the core of the Arches cluster and the surrounding
cloud region listed in Table~\ref{tab:regions}. The cloud region is
represented by an ellipse excluding the circular region of the cluster's
core.


\subsection{\nu}
\label{sec:data:nustar}

The {\it Nuclear Spectroscopic Telescope Array} (\nu) hard X-ray
orbital telescope \citep{nustar} provides arcminute angular resolution
imaging at energies above 10~keV not accessible by any previously or
currently operating missions. \nu\ carries two identical co-aligned
X-ray telescopes with an angular resolution of $18''$ (full width at
half maximum, FWHM). The focal plane detector units of each telescope,
referred to as focal plane module A and B (FPMA and FPMB), cover a
wide energy band $3-79$~keV, and provide spectral resolution of 400 eV
(FWHM) at 10 keV.

The Arches cluster was serendipitously observed during the Galactic
Center (GC) region campaign \citep{mori2015} with the \nu\ in 2012
October. Despite the fact that the Arches was observed at high
 off-axis angle, i.e. with low efficiency, K14 showed that the continuum
$10-20$~keV emission at that time was significantly detected around
the Arches cluster with a spatial morphology consistent with the Fe
K$\alpha$ fluorescent line emission. To better constrain the spectrum,
morphology and variability of the hard X-ray non-thermal emission of
the Arches cluster, we initiated the dedicated on-axis observations
performed in 2015 with the total exposure of 200~ks.

\begin{figure}
\includegraphics[width=\columnwidth]{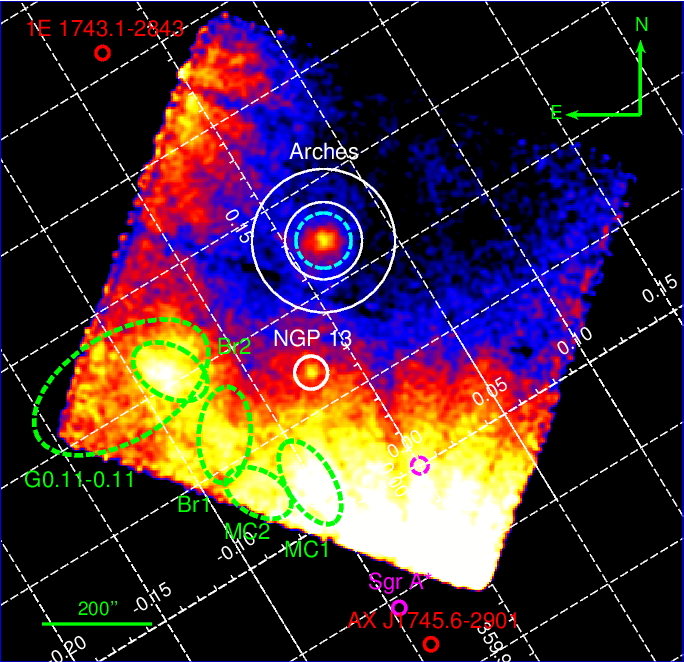}
\caption{\nu\ image of the Arches cluster region in $3-79$~keV energy
  band based on FPMA data only. The image is based on two \nu\
  observations and comprises 215~ks of exposure time. The position of
  the Arches cluster is shown by the cyan circle ($R=50''$) marked in
  a dashed style. The source and background spectrum are extracted,
  respectively, from the same green circle and annulus region ($R_{\rm
    min}=70''$, $R_{\rm max}=130''$). The sources of ghost-ray
  contamination are marked in red. Green sky regions reveal positions
  of selected Sgr A molecular clouds \citep{clavel2013}. NGP~13
  denotes position of an X-ray point source from the \nu\ survey of
  the Galactic center region \citep{hong2016}, also being a
  counterpart of the {\it Chandra} point source
  CXOUGC~J174551.9$-$285311 \citep{muno2009}.  Magenta circle
    shows the center of the Galactic
    coordinates.}\label{fig:nustar:arches}
\end{figure}

The \nu\ detectors can register X-rays passing outside the X-ray
optics modules due to gaps in detector shielding \citep[see
e.g.][]{wik2014,mori2015}. Unfocused flux of direct (``zero-bounce'')
photons or so called ``stray-light'' can be a significant contributor
to the detector background if there are bright X-ray sources within
$2-5$~deg of the \nu\ field of view (FOV). Examining FPMB data, we
noticed strong stray-light contamination from the bright X-ray source
GX~3+1 \citep{seifina2012} covering the Arches cluster region, which
led to rejection of FPMB data in the following analysis. In addition
to that FPMA data taken in 2015 were contaminated by ghost-rays (``one
bounce'' photons) from two nearby bright sources. The visible
ghost-ray patterns are consistent with the position of persistent
X-ray source 1E 1743.1-2843 \citep[see e.g.][and references
therein]{lotti2016} and transient LMXB AX~J1745.6-2901, which started
its prolonged outburst in 2013 July \citep{degenaar2015,ponti2015} and
faded in 2016 \citep{degenaar2016}. Nevertheless, the Arches cluster
core and cloud are not  strongly contaminated and can be well
localized in FPMA images  as obviously seen in
Fig.~\ref{fig:nustar:arches}, and also based on actual ghost-rays
  seen in other \nu\ observations \citep{bodaghee2014}.

Note that the celestial coordinates of each photon registered with
\nu\ are determined with some uncertainty caused by thermal bending
and external forces acting on the mast during orbit. The systematic
shift of the source position can be as high as $10''$. We noticed an
offset of the cluster core in 2015 data at the level of
$\sim4''$. Since the first \nu\ serendepitous observation of the
\arches\ taken in 2012 demonstrated good agreement between the
position of the cluster core and bright {\it Chandra} sources in it,
we use core centroid coordinates R.A.=17$^h$45$^m$50.52$^s$,
Dec.=$-28^{\circ}49'22.41''$ measured in K14 as a reference
position\footnote{Note that Table~3 in K14 contains mistaken centroid
  position of the Arches cluster core, the coordinates
  R.A.=17$^h$50$^m$50.43$^s$, Dec.=$-28^{\circ}49'23.07''$ should be
  read as R.A.=17$^h$45$^m$50.52$^s$, Dec.=$-28^{\circ}49'22.41''$.}
to correct sky coordinates of each incoming photon in 2015 data set.

The angular separation of the Arches cluster's core and cloud is
sufficient to spatially resolve these emission components with
\nu. However, spectral analysis of the given area is seriously
complicated due to the wide wings of the \nu\ PSF: the corresponding
half-power diameter (HPD, enclosing half of the focused X-rays),
reaches $\sim60''$ \citep{madsen2015}, which causes partial confusion
of the Arches cluster and its cloud. Taking the above into account,
we extract the spectrum of the Arches cluster from a circular region with
$R=50''$ pointed at the cluster centroid position, covering both the
cluster core and the cloud. The spectrum was extracted using the {\it
  nuproducts} task of the the \nu\ Data Analysis Software ({\sc
  nustardas}) v.1.5.1 and {\sc heasoft} v6.17. The background spectrum
was extracted from the annulus region within the radii range
$R=70-130''$, chosen as a result of a trade-off between the
representative background and ghost-ray contamination.

\section{Spectral modeling}
\label{sec:spec}

In this section we describe the details of our spectral analysis of
the different regions of the Arches cluster, namely: i) the core of
the stellar cluster within $R=15''$ circular region; ii) the cloud
ellipse region; and iii) the circular area of R=$50''$ containing both
cluster core and cloud, also referred later as `the Arches cluster
complex' (for the region coordinates see Table~\ref{tab:regions}). We
first analyse \xmm\ and \nu\ data alone, comparing with previous
observations and constructing the time evolution of the cloud
non-thermal emission, then we present the first \xmm\ and \nu\
broad-band joint spectral fit of the Arches cluster complex, and
finally perform 2D image analysis to separate stellar cluster and
cloud emission.  The uncertainty estimates of the fitted model
parameters in the following discussion are shown at the $90\%$
confidence level.

We begin with spectral analysis of the Arches core emission to check
the consistency of its spectral shape with previous observations and
to validate our data reduction procedures.

\subsection{\xmm: Cluster emission}
\label{sec:spec:xmm:cluster}

\begin{figure}
\includegraphics[width=\columnwidth]{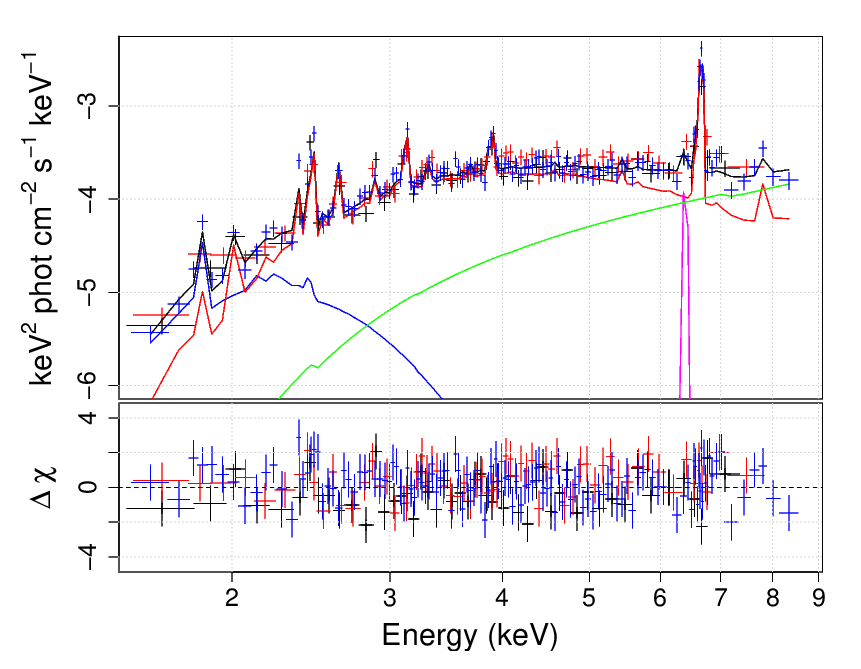}
\caption{X-ray spectrum of the Arches cluster based on the long \xmm\
  observation performed in 2015. The spectrum was extracted from
  $R=15''$ circular region (Table~\ref{tab:regions}). MOS1, MOS2, and
  PN data are shown in black, red, and blue colors, respectively. The
  fitted model includes two \apec\ components with
  $kT=0.22_{-0.05}^{+0.08}$~keV (blue) and
  $kT=1.58_{-0.09}^{+0.13}$~keV (red), and non-thermal power-law
  continuum (green) with Gaussian 6.4~keV line coloured in magenta
  (Model~2 in Table~\ref{tab:spe:cluster}). Hereafter, all spectra are
  shown in logarithmic scale, with Y axis labelled with only the
  exponents.}\label{fig:spe:cluster}
\end{figure}

We extracted the spectrum of the Arches cluster from a $15''$ circular
region and the corresponding instrumental and astrophysical background
from an annulus region around the cluster
(Table~\ref{tab:regions}). The Arches cluster X-ray emission, which
likely originates in one or more extreme colliding wind massive star
binaries \citep{capelli11a}, is dominated by a thermal emission with
prominent iron 6.7~keV line, however, at the same time, contribution
from X-ray bright clouds is not negligible. We thus model the emission
of the cluster region with an \apec\ plasma component and a
non-thermal component composed by a power-law continuum and Fe 6.4~keV
line. All the spectral emission components were subject to
interstellar photo-electric absorption modelled with {\it wabs} in
{\sc xspec}. \apec\ thermal plasma model is characterized by $kT$,
Z/Z$_{\sun}$, and I$_{kT}$ parameters, respectively, describing
temperature, metallicity relative to solar, and normalization in units
of $10^{-18}\int n_{e}n_{H}dV/(4\pi D^{2})$, where $n_e$ and $n_H$ are
the electron and proton number densities in units of
$\textrm{cm}^{-3}$, and D is the cluster distance in cm. Based on the
previous observations of the Arches cluster with \xmm\
\citep[T12,][]{wang2006} we fixed the metallicity to $Z=1.7 Z\sun$
throughout the paper. The best-fitting parameters of the current model
are summarized in Table~\ref{tab:spe:cluster} and the corresponding
MOS1/MOS2/PN spectra are shown in Fig.~\ref{fig:spe:cluster}. Note,
that we also fixed the energy of the 6.4~keV line due to its low flux
relative to the bright 6.7~keV line, which cause systematic shift
towards higher energies.

The value of $N_{\rm H}$ is higher than the Galactic value in the
direction to the Arches cluster ($N_{\rm H}=1.2\times10^{22}$
cm$^{-2}$; \citealt{2005A&A...440..775K}), by an order of magnitude,
which indicates strong local absorption, most likely caused both by
strong stellar winds of massive stars in the cluster and nearby
molecular material of the cloud. The temperature of the thermal
emission $kT=1.82\pm0.10$\,keV is consistent with the previous
measurement \citep[T12,][]{capelli11a}. The hard power-law continuum
$\Gamma\sim 1$ emission is also in agreement with that found in
previous studies.  However, the uncertainties, mainly caused by the
low statistics and the soft energy range $E<10$~keV, are
large. \cite{capelli11a} reported on 70\% brightening of the Arches
cluster emission in March/April 2007 relative to an average level of
$F_{\rm 2-10\ keV}=1.5\pm0.1\times10^{-12}$\,\ergscm\ ($2-10$~keV,
absorption corrected). Using the {\it cflux} model in {\sc xspec} we
estimated a $2-10$~keV unabsorbed total flux of the cluster of $F_{\rm
  2-10\ keV}=1.50_{-0.25}^{+0.30}\times10^{-12}$\,\ergscm, which is
fully consistent with the normal state flux level reported by
\cite{capelli11a}.

The second \apec\ component emission ($kT\sim0.2$~keV) which could be
due to a collection of individual stars, as suggested by T12,
significantly improves the fit statistics (expressed by reduced
$\chi^{2}_{\rm r}=1.05$ for 238 degrees of freedom, d.o.f.). In
contrast to T12, introducing separate absorber for the soft component
of the cluster emission did not provide an improvement to the fit
($\chi^{2}_{\rm r}$/d.o.f. = $1.15/237$), which is most likely caused
by the low sensitivity to the soft spectral component of our data set,
based on lower exposure compared to that used by T12.

\begin{table}
\noindent
\centering
\caption{Best-fitting spectral model parameters for the Arches cluster
  core emission ($R=15''$ circular region) measured with
  \xmm. }\label{tab:spe:cluster}
\centering
\vspace{1mm}
 \begin{tabular}{|c|c|r|r|r|c|c|}
\hline\hline
Parameter & Unit &  Model 1$^{*}$ &  Model 2$^{**}$    \\
\hline
$N_{\rm H}$ & $10^{22}$~cm$^{-2}$ & $8.73\pm0.50$ & $11.23\pm1.00$  \\ 
\hline
$kT$ & keV &  & $0.22_{-0.05}^{+0.08}$  \\
$I_{\rm kT}$ & (see Sect.~\ref{sec:spec:xmm:cluster}) &  &  $1.91_{-1.62}^{+16.61}\times10^{3}$  \\

\hline
$kT$ & keV & $1.82\pm0.10$ & $1.58_{-0.09}^{+0.13}$  \\
$I_{\rm kT}$ & (see Sect.~\ref{sec:spec:xmm:cluster}) & $13.60\pm2.00$ & $22.54\pm5.00$  \\
\hline
$E_{\rm 6.4\ keV}$ & keV &  6.4 (fixed) & 6.4 (fixed) \\
$N_{\rm 6.4\ keV}$ & $10^{-7}$ ph~cm$^{-2}$~s$^{-1}$ & $1.82_{-1.82}^{+3.47}$ &
$1.55_{-1.55}^{+3.62}$ \\
\hline
$\Gamma$& & $1.13\pm0.70$ & $0.33\pm1.00$ \\
$I_{\rm pow}$& 10$^{-5}$~cm$^{-2}$~s$^{-1}$~keV$^{-1}$ &
$1.86_{-1.42}^{+4.07}$ & $0.38_{-0.34}^{+1.93}$ \\
\hline
$\chi^{2}_{\rm r}$/d.o.f.& & 1.13/240  & 1.05/238  \\

\hline
\end{tabular}\\
\begin{flushleft}
*) WABS $\times$ (APEC + Gaussian + power law); \\
**) WABS $\times$ (APEC + APEC+ Gaussian + power law); 
\end{flushleft}
\vspace{3mm}
\end{table}

\subsection{\xmm: Cloud emission}
\label{sec:spec:xmm:cloud}

In this section we describe the spectral analysis of the cloud
emission based on the data taken during the long \xmm\ observation on
September 27, 2015. The main question we address here is the evolution
of the cloud non-thermal emission since 2012, when a drop was detected
as reported by C14. Strong variation implies that a large fraction of
the non-thermal emission of the cloud is due to the reflection of an
X-ray transient source.

To track the evolution of the cloud emission in a way that is
consistent with previous works (T12, C14), we extract \xmm\ spectrum
from $25''\times 59''$ ellipse region excluding the cluster emission
(Table~\ref{tab:regions}). Folowing C14 we consider the $2-7.5$~keV
energy range and spectral model $wabs \times (apec + powerlaw) +
gaussian$, fixing the temperature of the {\it apec} plasma component
to the $kT=2.2$~keV, metallicity to $Z=1.7 Z\sun$, the power-law index
to $\Gamma=1.6$, the centroid energy and width of the gaussian to
$6.4$~keV and $10$~eV, respectively (C14). Due to the low statistics
of the cloud region and the limited energy range selected
($2-7.5$~keV), the $2.2$~keV {\it apec} component represents an
average temperature plasma at the position of the cloud, accounting
for the soft and hard thermal diffuse emission components at 1 and 7
keV, usually considered to model the thermal background at the GC
\citep[e.g.][]{muno2004}. For this specific analysis, we used the
\xmm\ instrumental background as described in C14.  The best-fitting
parameters of this model obtained in the $2-7.5$~keV energy interval
are listed in Table~\ref{tab:spe:cloud} (Model 1). Note that the fit
is relatively poor ($\chi^{2}_{\rm r}$/d.o.f.= 1.54/224), mainly due
to the strong line excess at $\sim2.45$~keV, presumably belonging to
K$\alpha$ line from He-like Sulfur (S). We first tried to add a
low-temperature thermal emission component, however this approach: i)
did not remove the line at $\sim2.45$~keV, and hence ii) did not
improve the fit statistics; and iii) introduced the significant
deviation relative to C14 method, making it difficult to compare with
the long-term evolution of the cloud observed by C14. Therefore, we
follow an `ad-hoc' approach described as follows. Since the continuum
of low thermal emission was already taken into account by an average
$2.2$~keV {\it apec} plasma component, we extended the spectral model
by adding a Gaussian line at $\sim2.45$~keV (Model~2 in
Table~\ref{tab:spe:cloud}). Model~2 provides a better fit statistics
($\chi^{2}_{\rm r}$/d.o.f.=1.21/222) and it does not significantly
change the parameters of the non-thermal component
(Table~\ref{tab:spe:cloud}). This allows us to directly compare with
the C14 lightcurve of the Arches cloud both in the 6.4~keV line
emission and the power-law continuum.

C14 pointed out that the overall absorption and normalization of the thermal
plasma are compatible with being constant over the 13-year (2001-2014)
period with the estimated weighted mean values of $N_{\rm H}= 6.0 \pm 0.3
\times 10^{22}$~cm$^{-2}$ and $I_{\rm 2.2\ keV} = 3.6 \pm 0.7 \times
10^{-4}$~cm$^{-5}$, respectively. Our fitting procedure gives a 
comparable absorption and a marginally higher plasma normalization
$I_{\rm 2.2\ keV}$.

The equivalent width of the Fe $6.4$~keV emission line was calculated
with respect to the power-law continuum using the task {\it eqwidth}
in {\sc sherpa} package \citep{sherpa}, a part of the {\sc ciao-4.7}
software \citep{ciao}. C14 showed that the equivalent width of the Fe
K$\alpha$ line is compatible with being constant over time with an
average value $EW = 0.9 \pm 0.1$~keV \citep[also consistent with other
studies][T12, K14]{capelli11b}, indicating that most of the power-law
continuum is linked to the 6.4~keV emission. Our results obtained from
the new 2015 \xmm\ observations show that the line strength decreased
to $EW = 0.6-0.7$~keV (see Table~\ref{tab:spe:cloud}).

\begin{figure}
\includegraphics[width=\columnwidth]{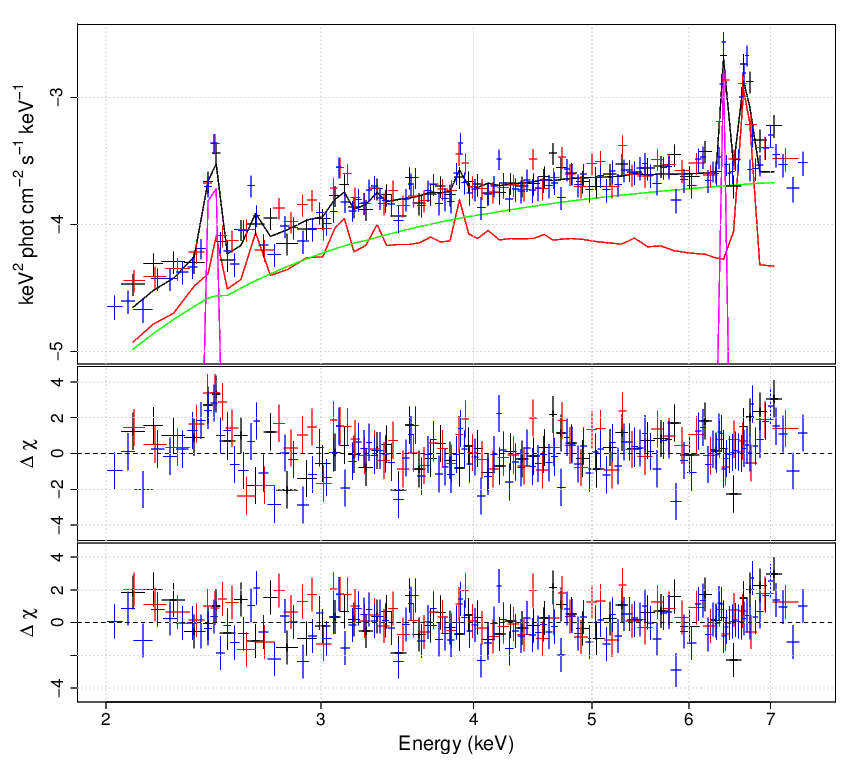}
\caption{ X-ray spectrum of the Arches cloud (cluster excluded)
  emission obtained with \xmm\ MOS1 (black), MOS2 (red) and PN (blue)
  cameras in 2015. The model spectrum is represented by an absorbed
  thermal \apec\ component ($kT=2.2$~keV, in red), non-thermal
  power-law continuum ($\Gamma=1.6$, in green) with 6.4~keV fluorescent
  line emission (magenta), and K$\alpha$ line of He-like S at 2.45~keV
  (magenta), see Sect.~\ref{sec:spec:xmm:cloud} and
  Table~\ref{tab:spe:cloud} for details. Two bottom panels show
  residuals (data minus folded model) in terms of sigmas with error
  bars of size one. The middle and bottom panel demonstrate residuals
  of the model without and with S K$\alpha$ line included,
  respectively.}\label{fig:spe:cloud}
\end{figure}

\begin{figure}
\includegraphics[width=\columnwidth]{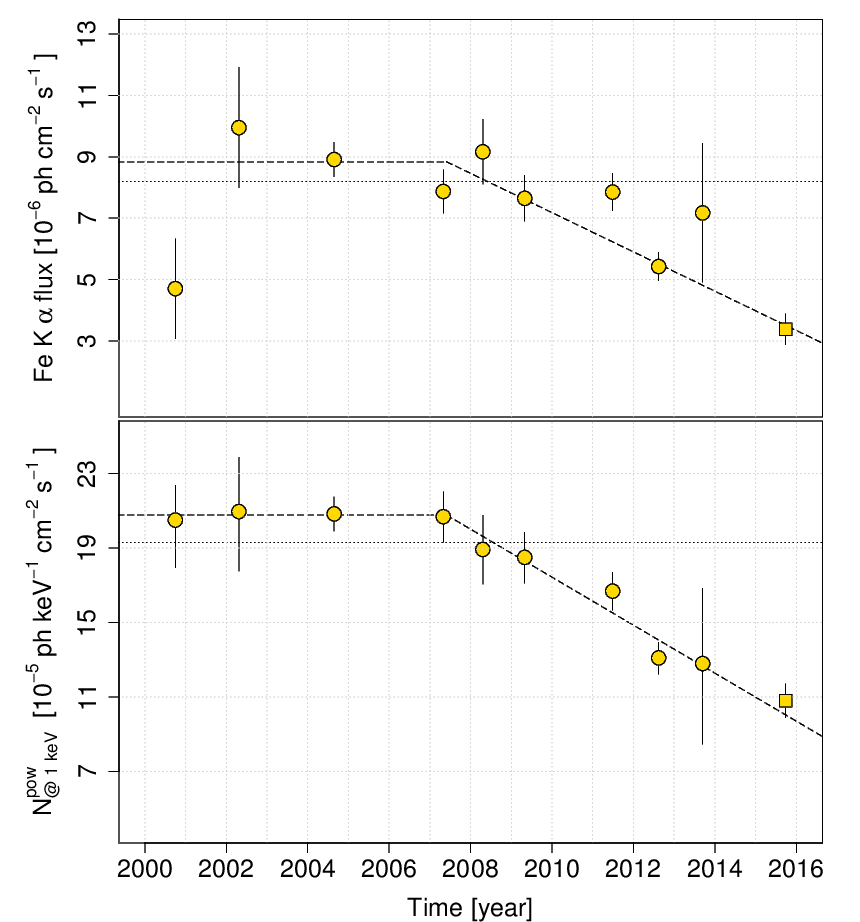}
\caption{The evolution of the non-thermal emission of the Arches cloud
  as traced by Fe K$\alpha$ line flux (upper panel) and the reflection
  continuum (bottom panel) represented by a power-law. The declining
  trend of the cloud emission reported by C14 (circle points) is
  confirmed by the recent measurements of the current work (square
  points). The dotted line represents an average emission evaluated up
  to 2011 by C14. The fit of the constant emission followed by a
  linear declining trend is shown by a dashed line.}\label{fig:clavel:track}
\end{figure}

The overall evolution of the normalizations of the 6.4~keV line 
and the reflection continuum is shown in Fig.~\ref{fig:clavel:track},
where the declining trend of the non-thermal emission of the Arches
cloud is clearly seen. Comparing the current normalizations of the
Fe~6.4~keV line and the power-law continuum with the average
emission up to 2011 as reported by C14, $N^{\rm 6.4\ keV}=8.2
\times 10^{-6}$~ph~cm$^{-2}$~s$^{-1}$ and $I_{\rm @\ 1\ keV}^{\rm pow}=19.3
\times 10^{-5}$~ph~cm$^{-2}$~s$^{-1}$~keV$^{-1}$, we obtain 
declining factors of $\sim2.4$ and $\sim1.8$, respectively.

\begin{table*}
\noindent
\centering
\caption{The list of the best-fitting parameters for the constant emission
   flux $N$ and linear law $C+\alpha\times T$ after $T_{\rm break}$
   describing the evolution of the power-law non-thermal continuum and Fe
   K$\alpha$ flux of the Arches cloud (see Sect.~\ref{sec:spec:xmm:cloud}).}\label{tab:evo}
 \centering
 \vspace{1mm}
 \begin{tabular}{|c|c|c|c|c|c|c|}
\hline\hline
Parameter & Units & \multicolumn{2}{c|}{Value}  \\
\hline
                &         & power-law & Fe K$\alpha$ line \\

$N^{*}$ & phot~keV$^{-1}$~cm$^{-2}$~s$^{-1}$ &
$20.78\pm0.71\times10^{-5}$ & $8.84\pm0.34\times10^{-6}$ \\
$\alpha$ &
phot~keV$^{-1}$~cm$^{-2}$~s$^{-1}$~year$^{-1}$ &
$-1.29\pm0.20\times10^{-5}$ &  $-0.64\pm0.07\times10^{-6}$ \\
$T_{\rm break}$ &  year &  $2007.41\pm0.51$ & 2007.41 (fixed) \\

\hline
\end{tabular}\\
\begin{flushleft}
*) The normalization of the power-law is defined at 1~keV; \\
\end{flushleft}
\vspace{3mm}
\end{table*}

The 2015 \xmm\ data allow us to better describe the observed trend of
the Arches cloud non-thermal emission. As a zero-order approach, we
model the observed flux as a constant power-law normalization
parameter $N^{\rm pow}$ followed by a linear law as a function of time
$C+\alpha\times T$ after some moment of time $T_{\rm break}$ ($C$
constant is derived from the continuity condition). Due to better
statistics the continuum flux is better determined from the data
relative to 6.4~keV line flux. The estimated best-fitting parameters
for the continuum flux evolution is listed in Table~\ref{tab:evo} and
the evaluated model light curve is shown in
Fig.~\ref{fig:clavel:track} (lower panel). Assuming that the 6.4~keV
line flux started to decline at the same time as the continuum, we run
the fitting procedure with frozen $T_{\rm break}$ parameter for the Fe
K$\alpha$ line flux light curve as shown in the upper panel of
Fig.~\ref{fig:clavel:track}. Applying the ratio between the flux
normalizations $N^{\rm 6.4\ keV} / N^{\rm pow} =
(4.25\pm0.22)\times10^{-2}$~keV to the declining scale of the
continuum $\alpha^{\rm pow}$ we obtain the corresponding estimate
$\alpha^{\rm 6.4\
  keV}=(-0.55\pm0.09)\times10^{-6}$~phot~cm$^{-2}$~s$^{-1}$~year$^{-1}$
statistically consistent with that independently determined from the
fit. We conclude that most of the power-law continuum was linked to
the 6.4 keV line emission on a long time scale before 2015. This
confirms the result obtained by C14 based on the correlation
analysis. However we should note that the recent change of the $EW$ in
2015 points towards an on-going decoupling of the continuum and line
emission.

The corresponding decay time, when the flux decreases by a factor of
two, was estimated to be $\tau_{1/2}=8.05\pm1.28$~years.  Based on the
sensitivity of the current \xmm\ data set and assuming constant fading
rate we would expect the non-thermal emission from the Arches cluster
to fall below the $2-10$~keV detection limit of
$2.8\times10^{-14}$~erg~cm$^{-2}$~s$^{-1}$ for 100~ks exposure after
2023.

\begin{table}
\noindent
\centering
\caption{Best-fitting parameters for two models applied to \xmm\
  spectrum of the Arches cloud (ellipse $25''\times 59''$, cluster excluded). Model~1 is identical to that used in
  C14; Model~2 contains extra Gaussian line at 2.45~keV.}\label{tab:spe:cloud} 
\centering
\vspace{1mm}
  \begin{tabular}{|c|c|r|c|c|c|c|}
\hline\hline
Parameter & Unit & Model 1$^{*}$ & Model 2$^{**}$ \\
\hline
$N_{\rm H}$ & $10^{22}$~cm$^{-2}$ & $6.51\pm0.34$ & $7.08\pm0.40$ \\
\hline
$kT$ & keV & 2.2 (fixed) & 2.2 (fixed) \\
$I_{\rm kT}$ & (see Sect.~\ref{sec:spec:xmm:cluster}) & $4.78\pm0.66$ & $4.46\pm0.71$ \\
\hline
 $E^{\rm 2.44\ keV}$ & keV &  & $2.445\pm0.011$ \\
 $N^{\rm 2.45\ keV}$ & $10^{-6}$ ph~cm$^{-2}$~s$^{-1}$ &  & $12.09\pm3.00$\\
 $EW^{\rm 2.45\ keV}$ & eV & & $130\pm12$ \\
\hline
 $E^{\rm 6.4\ keV}$ & keV & 6.4 (fixed) & 6.4 (fixed) \\
 $N^{\rm 6.4\ keV}$ & $10^{-6}$ ph~cm$^{-2}$~s$^{-1}$ & $3.06\pm0.43$ & $3.38\pm0.49$\\
 $EW^{\rm 6.4\ keV}$ & eV & $650\pm60$ & $608\pm57$ \\
\hline
 $\Gamma$ & & 1.6 (fixed)  & 1.6 (fixed) \\
 $N^{\rm pow}_{\rm @keV}$ &
 $10^{-5}$~ph~keV$^{-1}$~cm$^{-2}$~s$^{-1}$ & $10.19\pm0.80$ & $10.79\pm0.90$ \\
\hline
$\chi^{2}_{\rm r}$/d.o.f. & &  1.54/224 & 1.21/222 \\
\hline
\end{tabular}\\
\begin{flushleft}
*) WABS $\times$ (APEC + power law) + Gaussian; \\
**) WABS $\times$ (Gaussian + APEC + power law) + Gaussian; 
\end{flushleft}
\vspace{3mm}
\end{table}

\subsection{\nu: Cluster and Cloud X-ray emission}
\label{sec:spec:nustar:arches}

In this section we describe the spectral analysis of the Arches
cluster spectrum based on dedicated \nu\ observations performed with
the target on-axis during two 100~ks pointings in 2015
(Table~\ref{tab:log}). We check the consistency of the spectral shape
of the Arches cluster with previous low-efficiency observations in
2012 when it was detected at large off-axis angles (K14). As
mentioned in Sect.~\ref{sec:data:nustar} we utilize FPMA data only,
extracting the source spectrum from a circular region with a radius
of $50''$ centred at the cluster's core and including the cluster as
well as the cloud emission. The background spectrum was measured in an
annular region (Table~\ref{tab:regions}).

\begin{figure}
\includegraphics[width=\columnwidth]{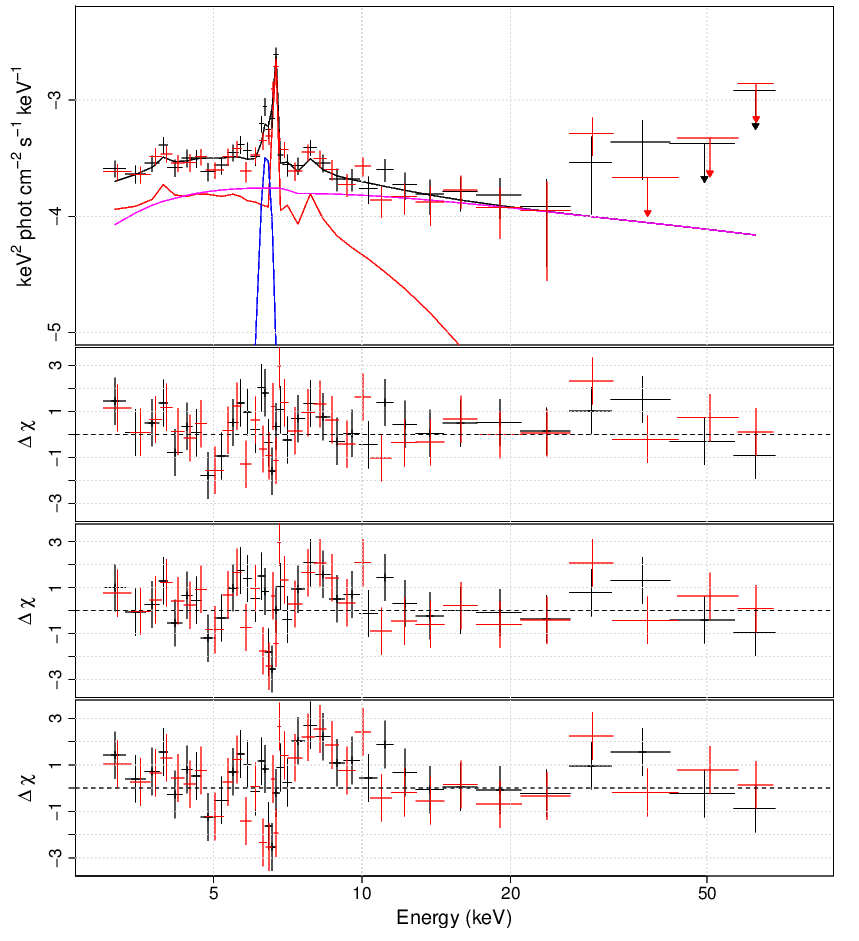}
\caption{{\it Upper panel:} X-ray spectrum of the Arches cluster
  region as measured with FPMA during the 1st and 2nd \nu\
  observations, shown, respectively, in red and black data points. The
  lines represent best-fitting spectral model (in black) with \apec\ (red),
  power-law (magenta) and 6.4~keV Gaussian line (blue) components
  (Model 1 in Table~\ref{tab:spe:nustar:arches}). Given the
  cross-normalization factor $C$ between the observations close to
  unity, only one model is shown for simplicity (1st
  observation). {\it Bottom panels} show corresponding residuals for
  three models listed in Table~\ref{tab:spe:nustar:arches}: power law
  with 6.4 keV Gaussian line, LECR{\it p}, and {\it
    REFLIONX}.}\label{fig:spe:nustar:arches}
\end{figure}

To be consistent with K14 analysis we utilize three spectral models,
each containing a collisionally ionized plasma emission model (\apec)
representing the cluster's core thermal emission. A non-thermal
emission component contains the simple power-law with a Gaussian line at
6.4~keV or cosmic-ray (CR) induced emission model or a self-consistent
X-ray reflection model. Despite LECR-only emission model is almost excluded
based on variability, we continue to test it because steady background
level could be a result of CR heating, while most of the varying
emission is due to reflection. All the emission components were
subject to a line-of-sight photoelectric absorption model {\it wabs}
in {\sc xspec}. As the energy response of \nu\ is not very sensitive
for measuring the absorption column, we fixed the total absorption
column at $N_{\rm H}=9.5\times 10^{22}$~cm$^{-2}$ measured by T12 in
the core of the Arches cluster. The best-fitting model parameters
obtained as a result of the fitting procedure are listed in
Table~\ref{tab:spe:nustar:arches}, while the \nu\ spectra with model
residuals are shown in Fig.~\ref{fig:spe:nustar:arches}.

This simple phenomenological model with power-law provides a reliable
fit to the data with $\chi^{2}_{\rm r}$/d.o.f.=$0.97/364$. The
temperature of the cluster's core emission was not well constrained
and found to have a relatively high value of
$kT=2.40_{-0.40}^{+1.50}$~keV with respect to
$kT=1.76_{-0.29}^{+0.36}$~keV measured on the \nu\ data taken in 2012
with the same model (K14). The non-thermal continuum is characterized
by a steeper power-law with $\Gamma=2.41_{-0.34}^{+0.58}$ compared to
$1.62\pm0.31$ in K14. The unabsorbed $3-20$~keV flux of the continuum
was estimated to be $F^{\rm pow}_{\rm 3-20\
  keV}=(0.56_{-0.09}^{+0.12})\times10^{-12}$\,\ergscm\ compared to K14
measurement $F^{\rm pow}_{\rm 3-20\
  keV}=(1.49_{-0.24}^{+0.26})\times10^{-12}$\,\ergscm\ in 2012. The
drop of the continuum flux with a factor of $\sim3$ in three years is
stronger than a drop with a factor of 2 in eight years expected from
the \xmm\ data (Sect.~\ref{sec:spec:xmm:cloud} and
\ref{sec:joint}). The fitting procedure determines the centroid
position of Fe 6.4~keV line at $E_{\rm 6.4\ keV}=6.34\pm1.12$~keV
along with relatively low equivalent width ($EW_{\rm 6.4\ keV}$) in a
range $0.25-0.56$~keV compared to $EW_{\rm 6.4\ keV}\sim1$~keV
measured in previous works (T12, K14), confirming the lower EW
measured with \xmm. We conclude that, despite the fading of the Arches
cloud emission, the non-thermal component is well constrained in 2015
observations, and still dominates the Arches cluster emission above
10~keV.

We then fitted the non-thermal emission of the cloud with the
cosmic-ray induced emission model developed by T12, where the
collisional ionization of the cloud by low-energy cosmic-ray (LECR)
protons (LECR{\it p}) reproduces the observed 6.4 keV fluorescence
line width. In contrast, the LECR {\it electron} model (LECR{\it e})
requires a metallicity $\gtrsim 3$ times the solar value to account
for the measured $EW_{\rm 6.4\ keV}\sim1$~keV, which makes the
LECR{\it e} model hardly compatible with the measured properties of
the Arches cloud emission (T12, K14). We consider only the LECR{\it p}
model in this work. The model depends on the LECR path length in the
ambient medium, $\Lambda$, the minimum energy of the CR particles
penetrating the cloud, $E_{\rm min}$, the power-law index of the CR
source energy spectrum, $s$, the model normalization $N_{\textrm{\tiny
    LECR}}$ and the metallicity of the X-ray emission region, $Z$. We
fixed $\Lambda$ and $E_{\textrm{\tiny min}}$ parameters of the LECR$p$
model according to T12, considering the slope $s$ and the
normalization $N_{\textrm{\tiny LECR}}$ as free parameters. The model
provides acceptable fit statistics ($\chi^{2}_{\rm
  r}$/d.o.f.=1.02/367). The power-law index $s=2.30\pm0.37$ is
generally higher than $s=1.65^{+0.59}_{-0.55}$ measured by K14 with
the \nu\ data in 2012, however more consistent with
$s=1.9^{+0.5}_{-0.6}$ estimated by T12 for the LECR{\it p} model with
the \xmm\ data. The model normalization $N_{\textrm{\tiny LECR}}
=3.18_{-0.62}^{+1.00}\times10^{-8}$~erg~cm$^{-2}$~s$^{-1}$ implies a
power injected by LECR protons of $\sim3\times10^{38}$\lum, which is a
factor of $\sim2$ less than the corresponding value measured by
K14. Such a decrease rules out this model as the only contribution to
the non-thermal emission up to 2012.

The third model includes the self-consistent X-ray reflection model
{\sc REFLIONX} which we used to describe the Arches cluster non-thermal
emission in K14.  {\sc REFLIONX} describes the reflected spectrum for
an optically thick atmosphere of constant density, illuminated by
radiation with a power-law spectrum \citep{reflionx}. The model
predicts fluorescence lines and continuum emission. Originally
developed for the surface of hot accretion disks in binary systems,
{\sc REFLIONX} can be well applied for cold material around the Arches
cluster (K14), by fixing the ionization parameter $\xi$ at the lowest
allowed value of $10$~erg~cm~$s^{-1}$. The model fits the data
relatively well ($\chi^{2}_{\rm r}$/d.o.f.=1.06/366), with a somewhat
steeper and less constrained photon index for the illuminating
power-law spectrum $\Gamma_{\textrm{\tiny RX}}=3.30_{-0.19}^{+\infty}$
compared to $\Gamma_{\textrm{\tiny RX}}=2.93\pm0.31$ obtained by K14,
which points to general softening of the spectrum confirmed by
power-law model considered above.

We conclude that, similar to previous observations of the Arches
cluster with \nu, the 2015 spectral shape above $\sim10$~keV is equally
well described by the X-ray photoionization and CR-induced emission
models, giving no direct evidence to determine the nature of
the emission mechanism. Based on this, the following joint \xmm\ and
\nu\ spectral analysis is done only for simple power-law model.

\begin{table*}
\noindent
\centering
\caption{Best-fit spectral model parameters for the Arches cluster $R=50''$ region emission measured with
  \nu.}\label{tab:spe:nustar:arches}
\centering
\vspace{1mm}
 \begin{tabular}{|c|c|r|r|r|c|c|}
\hline\hline
Parameter & Unit &  Model 1 $^{*}$ &  Model 2 $^{**}$  &  Model 3 $^{***}$   \\
\hline
$N_{\rm H}$ & $10^{22}$~cm$^{-2}$ & $9.5$ (fixed) &  $9.5$ (fixed) &  $9.5$ (fixed) \\ 

\hline
$kT$ & keV & $2.40_{-0.40}^{+1.50}$ & $1.89_{-0.15}^{+0.21}$ & $1.95_{-0.17}^{+0.32}$ \\
$I_{\rm kT}$ &  (see Sect.~\ref{sec:spec:xmm:cluster})  & $9.17\pm5.28$ & $17.46\pm3.50$ &  $16.55\pm3.00$ \\
\hline
$\Delta E_{\rm 6.4\ keV}$ & keV &  0.1 (fixed) & & \\
$E_{\rm 6.4\ keV}$ & keV &  $6.34\pm1.12$ & & \\
$F_{\rm 6.4\ keV}$ & $10^{-6}$ ph~cm$^{-2}$~s$^{-1}$ &  $2.60\pm1.47$ & & \\
$EW_{\rm 6.4\ keV}$ & eV & $450\pm150$ &  & \\
\hline
$\Gamma$& & $2.41_{-0.34}^{+0.58}$ & & \\
$F^{\rm pow}_{\rm 3-20\ keV}$& 10$^{-13}$~erg~cm$^{-2}$~s$^{-1}$ &
$5.58_{-0.91}^{+1.15}$  & & \\
\hline
$\Lambda$ & (H-atoms cm$^{-2}$) &  & $5\times10^{24}$ (fixed) & \\
$s$ &   &  & $2.30\pm0.37$ & \\
$E_{\textrm{\tiny min}}$ & (keV/n)  & & $10^{4}$ (fixed) &  \\
$N_{\textrm{\tiny LECR}}$ &
($10^{-8}\textrm{~erg~cm}^{-2}\textrm{~s}^{-1}$)  &  & $3.18_{-0.62}^{+1.00}$ &  \\

\hline
$\Gamma_{\textrm{\tiny RX}}$ &  &  &  & $3.30_{-0.19}^{+\infty}$ \\
$\xi$ & erg~cm~s$^{-1}$  &  &  & $10$ (fixed) \\
$I_{\textrm{\tiny RX}}$ & ($10^{-5}$) &  &  & $7.16_{-3.52}^{+1.01}$ \\
\hline
$\chi^{2}_{\rm r}$/d.o.f.& & 0.97/364  &  1.02/367   &  1.06/366  \\

\hline
\end{tabular}\\
\begin{flushleft}
$^{*}$ Model 1: C $\times$ WABS $\times$ (APEC + Gaussian + power law); \\
$^{**}$ Model 2: C $\times$ WABS $\times$ (APEC + LECR{\it p}); \\
$^{***}$ Model 3: C $\times$ WABS $\times$ (APEC + REFLIONX); \\
\end{flushleft}
\vspace{3mm}
\end{table*}

\subsection{\nu\ and \xmm\ joint fit}
\label{sec:joint}

In the previous sections we analysed individual \xmm\ and \nu\ fits to
check the consistency of the spectral shape of the Arches cluster with
previous measurements, and we confirmed the decrease of the net flux of
the non-thermal emission of the surrounding cloud both in continuum
and Fe K$\alpha$ flux. Additionally, we detected a significant decrease
of the $EW$ of the 6.4~keV line, pointing to a dramatic change of
the non-thermal emission of the Arches cloud, also reflected in
morphology transformation. To extract more constrains on the spectral
shape of the Arches cluster emission (star cluster core and
extended power-law non-thermal emission together) we present the first
simultaneous \xmm+\nu\ fit on the Arches cluster complex.

\begin{figure}
\includegraphics[width=\columnwidth]{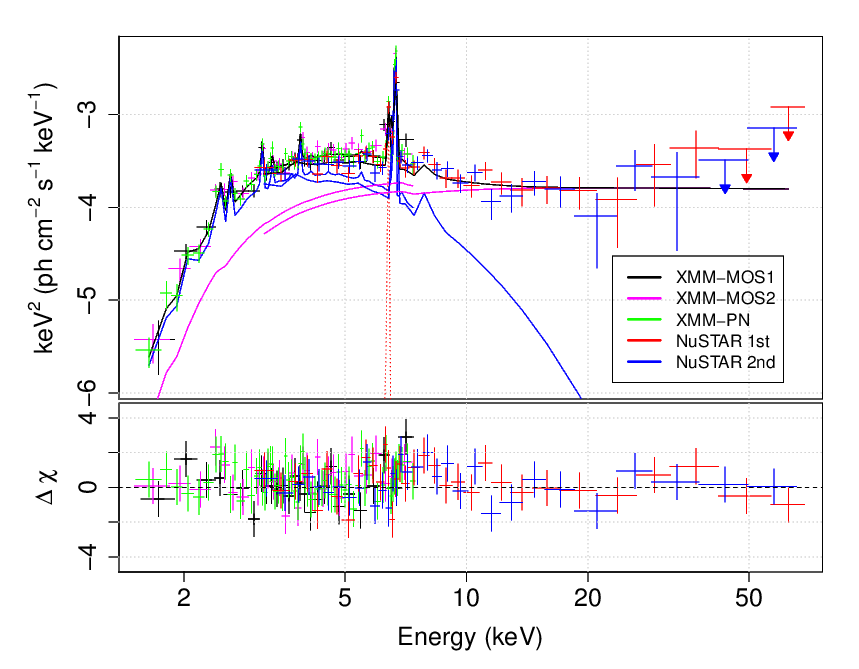}
\caption{Joint \xmm\ (MOS1, MOS2, PN) and \nu\ (FPMA) spectrum of the
  Arches cluster region extracted from $R=50''$ circle shown in
  Fig.~\ref{fig:nustar:arches}. Black line represents best-fit model
  (Table~\ref{tab:spe:joint:arches}), comprising thermal plasma
  (blue), fluorescent Fe K$\alpha$ 6.4~keV line emission (red) and
  non-thermal power-law continuum (magenta).}\label{fig:spe:joint}
\end{figure}

Fig.~\ref{fig:spe:joint} shows a joint \xmm\ (MOS1, MOS2 and PN) and
\nu\ spectrum of the Arches cluster region extracted from a $R=50''$
circular region positioned at cluster's centroid position, and the
corresponding background spectrum for each instrument, containing the
instrumental and astrophysical components, was extracted from the
annular region around the Arches cluster (see Table~\ref{tab:regions}
and Fig.~\ref{fig:nustar:arches}). The way the source region is
defined, the joint spectrum contains both thermal emission from the
cluster and non-thermal emission from the surrounding cloud.  To
describe the data, we use the model containing, respectively, an
\apec\ plasma model and a power-law continuum with a Gaussian line at
6.4~keV. We also added a cross-normalization constant between the
\xmm\ and \nu\ data, fixing the former to unity. The joint spectrum is
well described by the model with $\chi^{2}_{\rm
  r}$/d.o.f. =$1.00/918$.  The model parameters are listed in
Table~\ref{tab:spe:joint:arches}. The soft thermal component is not
required by the fit. The best-fit cluster temperature was found at
$kT=1.95\pm0.14$\,keV, which is in general agreement with \xmm\ and
\nu\ data analysed separately in Sect.~\ref{sec:spec:xmm:cluster} and
Sect.~\ref{sec:spec:nustar:arches}, respectively. The
cross-normalization parameter $C=0.82\pm0.04$ is somewhat lower than
that expected from relative \xmm\ and \nu\ cross-calibration results
\citep{madsen2015}, which is most likely caused by the overly
simplified background approximation of the \nu\ data, which is
contaminated by ghost-rays (Fig.~\ref{fig:nustar:arches}). We tested
spectral extraction with different ghost-ray-free background region,
and found strong deviations in the normalization of the \nu\
spectrum. For instance, using the faintest possible background region,
the normalisation of the \nu\ spectrum was a factor $\sim2$
higher. However, these ghost-ray free background regions also
introduce inaccurate deviations in the shape of the \nu\ Arches
cluster. Therefore, we conclude that the current annulus background
region is the best possible solution, even if it is slightly
overestimating the background contribution.

As seen from Table~\ref{tab:spe:joint:arches}, we confirm that the
$N_{\rm H}$ value chosen in the \nu\ spectral analysis
(Table~\ref{tab:spe:nustar:arches}), which was also used in K14, was
valid. The joint fit gives the first solid measurement of the
power-law $\Gamma$ up to $30-40$~keV and it is consistent with other
molecular clouds in the CMZ \citep[also measured by
\nu,][]{shuo2015,mori2015}, pointing towards a similar origin for all
clouds.

\begin{table}
\noindent
\centering
\caption{Best-fit spectral model parameters for the Arches cluster $R=50''$ region emission measured with
  \xmm\ and \nu. The model is described in {\sc xspec} notation as WABS $\times$ (APEC + Gaussian + power law).}\label{tab:spe:joint:arches}
\centering
\vspace{1mm}
 \begin{tabular}{|c|c|r|r|r|c|c|}
\hline\hline
Parameter & Unit &  Value \\
\hline
$N_{\rm H}$ & $10^{22}$~cm$^{-2}$ & $9.32_{-0.51}^{+0.86}$ \\ 

\hline
$kT$ & keV & $1.95\pm0.14$ \\
$I_{\rm kT}$ & (see Sect.~\ref{sec:spec:xmm:cluster}) & $18.55\pm3.00$  \\

\hline
$\Delta E_{\rm 6.4\ keV}$ & keV &  0.01 (fixed)  \\
$E_{\rm 6.4\ keV}$ & keV &  $6.38\pm0.02$  \\
$F_{\rm 6.4\ keV}$ & $10^{-6}$ ph~cm$^{-2}$~s$^{-1}$ &  $3.12\pm0.65$  \\
$EW_{\rm 6.4\ keV}$ & eV & $704_{-95}^{+102}$  \\
\hline
$\Gamma$& & $2.03\pm0.16$  \\
$I_{\rm pow}$& 10$^{-5}$~cm$^{-2}$~s$^{-1}$~keV$^{-1}$ &
$21.91_{-6.90}^{+10.2}$ \\
\hline

$C$& & $0.82\pm0.04$ &  \\

$\chi^{2}_{\rm r}$/d.o.f.& & 1.00/918   \\

\hline
\end{tabular}\\

\vspace{3mm}
\end{table}

\subsection{\nu: spatially resolved spectrum}
\label{sec:nustar:2d}

The Arches cluster region exhibits both thermal and non-thermal
emission components, originating, respectively, in the cluster's core
and nearby molecular cloud region. Despite the fact that both emission
components are relatively well resolved spatially, the spectral
analysis with the high resolution of \xmm\ and {\it Chandra} still
has to account for the combination of thermal and non-thermal emission
components, most likely affected by projection effects. For
instance, T12 model the emission of the Arches cluster region as
the combination of an \apec\ plasma component and a non-thermal
component represented by a power-law continuum and a Gaussian line at
6.4 keV.

\begin{figure}
\includegraphics[width=\columnwidth]{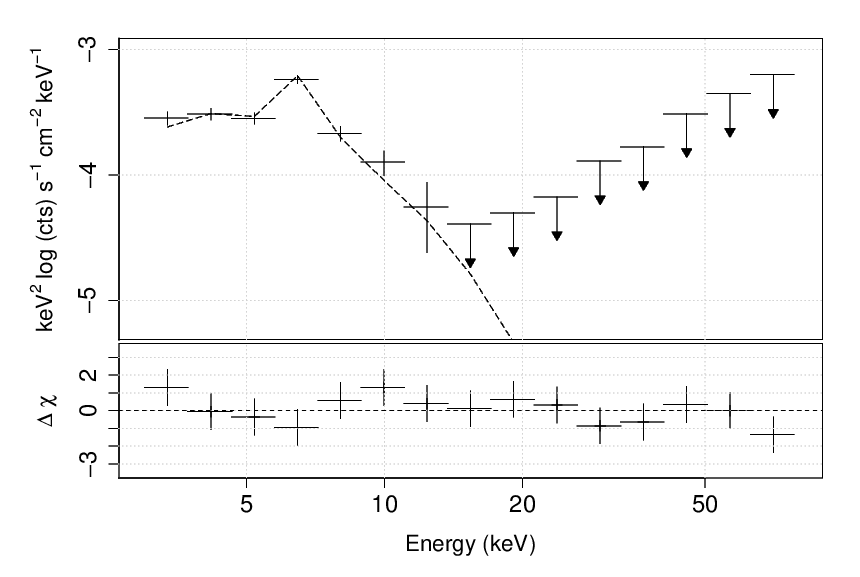}
\caption{Spatially resolved \nu\ spectrum of the Arches cluster core
  approximated with collisionally ionized plasma emission model \apec\
  with temperature $kT=2.44_{-0.26}^{+0.40}$~keV (dashed
  line).}\label{fig:nustar:2d:core}
\end{figure}

\begin{figure}
\includegraphics[width=\columnwidth]{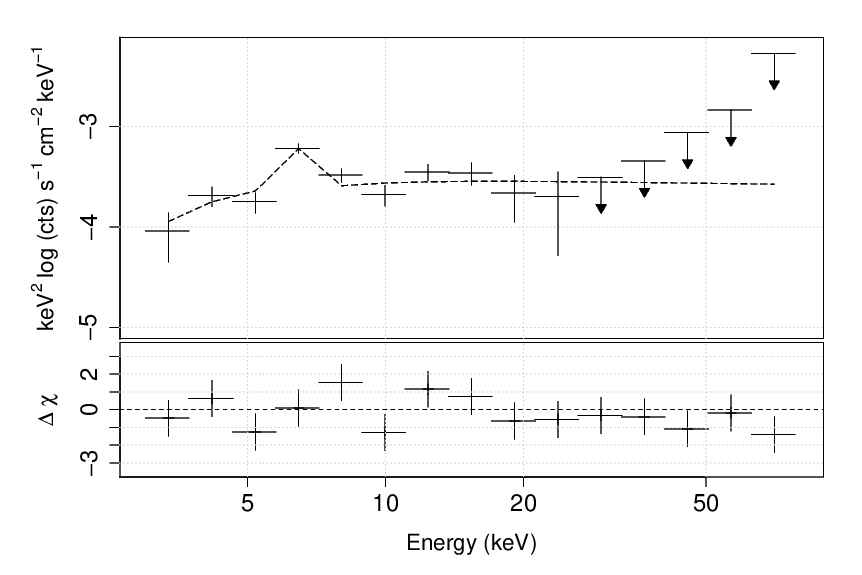}
\caption{Spatially resolved \nu\ spectrum of the Arches cloud modelled
  with power-law $\Gamma=2.06\pm0.26$ and 6.4~keV Gaussian
  line.}\label{fig:nustar:2d:cloud}
\end{figure}

Using the fact that thermal emission is concentrated in the cluster's
core, and non-thermal cloud emission is spread more widely (i.e. these
components are characterized by different spatial extent or frequency
as demonstrated with the use of wavelet analysis by K14), we attempt
to spatially decouple the cluster and cloud hard X-ray emission in the
\nu\ data in order to extract their X-ray spectra. The procedure is
described in Appendix~\ref{app:2d}, and the results are presented in
Fig.~\ref{fig:nustar:2d:core} and \ref{fig:nustar:2d:cloud}.

We fitted the cluster's core emission spectrum
(Fig.~\ref{fig:nustar:2d:core}) with one \apec\ model subjects to a
line-of-sight photoelectric absorption fixed at $N_{\rm H}=9.5\times
10^{22}$~cm$^{-2}$. We assume that this spectrum does not include the
extended non-thermal emission, so we do not need any additional
components in the spectrum. This simple model provides an acceptable
fit to the data with $\chi^{2}_{\rm r}$/d.o.f. =$0.66/13$. The only
two free parameters were estimated from the fit: the temperature of
the plasma $kT=2.44\pm0.40$~keV and the unabsorbed $3-8$~keV flux
$(8.70\pm0.70)\times10^{-13}$\,\ergscm. The cluster temperature is
marginally higher, but it is still consistent with $kT=1.6-2$~keV
measured with \xmm\ data, which is more sensitive to thermal emission
(Sect.~\ref{sec:spec:xmm:cluster}). In contrast, the $3-8$~keV
unabsorbed flux of the cluster's thermal component is somewhat lower
when compared to the corresponding value found with \xmm:
$(1.23\pm0.20)\times10^{-12}$\,\ergscm.

Based on the 2D image spectral analysis, we assume that the spatially
resolved spectrum of the Arches cloud contains the non-thermal
emission only, and it is not mixed with thermal radiation of the
cluster's core. The position and width of the Gaussian line was fixed to 6.4~keV
and 0.1~keV, respectively. The model gives acceptable fit statistics
$\chi^{2}_{\rm r}$/d.o.f.=$1.02/12$ and allows for a constraint on the power-law
slope $\Gamma=2.06\pm0.26$ (confirming the joint \xmm\ and \nu\ fit
results, see Sect.~\ref{sec:joint}) and the unabsorbed $3-8$~keV and $3-20$~keV
fluxes: $F_{\rm 3-8}=(4.92\pm1.00)\times10^{-13}$\,\ergscm\ and $F_{\rm
  3-20}=(9.33\pm1.34)\times10^{-13}$\,\ergscm. The total
flux of the 6.4~keV Gaussian line was estimated to be
$(1.38\pm0.50)\times10^{-5}$ photons~cm$^{-2}$~s$^{-1}$. Given the fit
statistics and uncertainties of the model parameters, we conclude that
addition of cluster's thermal emission is not required.

We then applied a more physically motivated X-ray reflection model
developed in the geometry of scattering X-ray incident photons from
\sgra\ to the observer (Appendix~\ref{app:chur}); however, we found
that quality of the data does not allow us to constrain the reflection
geometry.

\section{Discussion}
\label{sec:discussion}

During the long \nu\ and \xmm\ observations of the Arches cluster
region in 2015, dedicated to monitor the emission of the molecular
cloud, we observe the following distinct components in the X-ray
spectra of the cluster region: an optically thin thermal plasma with a
temperature $kT\sim1.6-2.0$~keV of the stellar cluster; an additional
thermal plasma component with $kT\sim0.18-0.20$~keV of the cluster;
and non-thermal emission of the cloud region characterized by a power-law
continuum and fluorescent Fe K$\alpha$ 6.4~keV line. In this work we
present the first solid measurement of the power-law $\Gamma\sim2$ up
to $30-40$~keV. The \nu\ spectrum extracted from a $50''$ circlular region
covering the cluster and the cloud region represents a mixture of the
components mentioned above. Using 2D image analysis, we decoupled
spectral components (Sect.~\ref{sec:nustar:2d}), and confirmed the
power-law slope $\Gamma$ of the non-thermal emission obtained with
regular spectral analysis. $\Gamma\sim2$ is consistent with other
molecular clouds in the CMZ \citep{shuo2015,mori2015}, which points
to a similar origin for all clouds.  In the reflection scenario, this
also means that the incident hard X-ray emission must have a
power-law spectrum with $\Gamma\sim2$ up to at least 40~keV, which
adds important evidence favouring the association with past activity
of \sgra, which exhibits a similar spectrum during X-ray flares
\citep{barriere2014}.

\begin{figure}
\includegraphics[width=0.99\columnwidth]{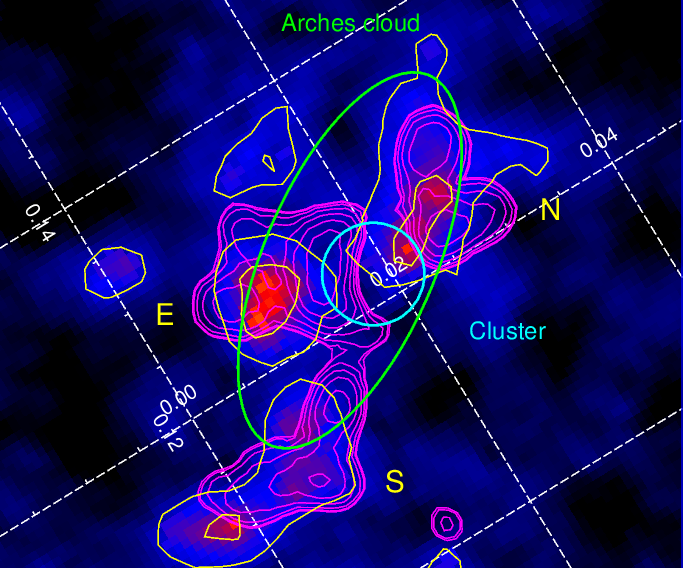}
\caption{\xmm\ K$\alpha$ line mosaic image of the Arches cluster
  region of Fe at 6.4~keV highlighted by yellow contours. Magenta
  contours show clumps of 6.4~keV emission based on low statistics map
  obtained in 2013 by C14.}\label{fig:xmm:arches:2013}
\end{figure}

The morphology of the non-thermal emission of the molecular clouds
near the Arches cluster reveals substructure with three bright knots
to the east (E), north (N), and south (S) of the cluster's core,
marked correspondigly in Fig.~\ref{fig:xmm:arches}. The current
6.4~keV line brightness distribution does not demonstrate a spatial
correlation with previous morphology studies made before 2012
\citep[][T12]{capelli11b}, which showed mainly two bright spots to the
north and to the south of the cluster (within the ellipse region). On
the other hand, we found a similar pattern in the 6.4~keV map shown by
C14 based on 2013 data set, which points towards a dramatic change of
the morhology since 2012. To illustrate this, we show the current
6.4~keV map in Fig.~\ref{fig:xmm:arches:2013} compared with contours
of bright 6.4~keV clumps observed in 2013 by C14. A similar changes in
morphology and appearence of new extended features are observed in
other parts of the Galactic Centre region \citep[e.g., Bridge, MC1,
MC2, G0.11-0.11 and Sgr~B2 molecular clouds, see
Appendix~\ref{app:SgrA}][]{ponti2010,clavel2013}. For instance,
\cite{shuo2015} revealed the substructure of the fading Sgr~B2 cloud,
with two compact cores, Sgr~B2(M) and Sgr~B2(N), and newly emerging
cloud G0.66$-$0.13, which highlights that the illuminating front(s)
propagating within this region has already started to leave its main
cores. The similitudes between the Arches cloud and the other
molecular clouds also favour a similar reflection mechanism at the
origin of their non-thermal emission. We confirm the variability seen
by \xmm, and measure, for the first time, variability of the
non-thermal emission of the Arches cloud with \nu, providing
additional evidence that continuum emission above 10~keV is linked to
the 6.4~keV line flux, which was not known a priori in this complex
region. Strong variations of the molecular clouds in the Galactic
Centre support the X-ray reflection mechanism of fluorescence observed
in CMZ and reveal propagation of illuminating fronts, presumably
induced by the past flaring activity of \sgra.

The similitude with other molecular clouds is also confirmed by the
long-term variability of the non-thermal emission of the Arches
cloud. The fluorescent Fe K$\alpha$ emission and hard X-ray continuum
continue to demonstrate fast time variability, as discovered by
C14. Thanks to the new observations with \xmm\ we are able to better
describe the declining trend of the cloud emission. The linear fit of
the varying flux, both in Fe K$\alpha$ line and hard X-ray continuum,
gives a half-time decay $\tau_{1/2} = 8.05 \pm 1.28$~years, with the
emission starting to decrease in $T_{\rm break} = 2007.41 \pm
0.51$. Such strong variations have been observed in several other
molecular clouds in the Sgr~A complex \citep[MC1, MC2 and
G0.11-0.11][]{capelli2012,clavel2013}, but also in the giant molecular
cloud Sgr B2 \citep{terrier2010}, which started to fade in 2001
\citep{inui2009} after a decade of possibly constant emission
\citep{revnivtsev2004a}. However, we should note that this tentative
comparison should be used with a caution: the declining trends for
both the Sgr~B2 region and the Arches cloud region are averaged over
relatively large regions, which could include sub-structures that may
vary differently \citep{shuo2015}.

Our analysis of \xmm\ and \nu\ data of the Arches cloud region
indicates a significant drop of the equivalent width of the Fe
K$\alpha$ line $EW_{\rm 6.4\ keV} = 0.6-0.7$~keV from the average
$EW_{\rm 6.4\ keV} = 0.9 \pm 0.1$~keV observed since 2002
\citep[][T12, K14, C14]{capelli11b}. The $EW$ decrease can trace a
relative increase of the LECR{\it p} contribution (which is known to
produce a lower $EW$). Indeed, while its absolute contribution is
expected to be constant over time, its relative contribution to the
total amount of non-thermal emission should be increasing as the
reflection component is decreasing. However, the drop in $EW$ could
also be due to a change in the geometry or the metallicity of the
reflector. Due to the dramatic recent changes in the morphology of the
6.4 keV emission, this second explanation (or a combination of the
two) should not be excluded.

In the X-ray reflection scenario, within a single cloud, the $EW$ is
only expected to increase with time \citep[based on absorption, see
e.g.][]{odaka2011}, so a lower $EW$ would mean that we are seeing a
structure that is at a different position along the line-of-sight. The
clumps of 6.4~keV line emission, visible in 2015 (and 2013), were not
seen to be bright in 2000-2012, which means that reflection is still
contributing to a certain extent to the faint 2015 emission and that
the illuminating front is now reaching different clumps. If these new
clumps are within the same cloud, this means that the illuminating
event is rather short (the light curve average over the 'cloud' region
is hiding faster variations) and if they are not part of the cloud,
they are likely to be in a very different location along the line of
sight and be illuminated by a different event.

\centerline{}

\section{Summary}
\label{sec:summary}

Recent \nu\ and \xmm\ observations of the molecular cloud around the
Arches cluster demonstrate a dramatic change both in morphology and
intensity of the non-thermal emission. Below we summarize the main
results of this work.

\begin{itemize}

\item Despite the observed fading, the non-thermal continuum emission
  of the cloud is detected up to $\sim30-40$~keV. The continuum is
  well approximated by a power-law with $\Gamma\sim2$, which is the
  first direct measurement of the Arches hard X-ray continuum,
  consistent with other molecular clouds measured with \nu\
  \citep{shuo2015,mori2015}

\item The relatively homogeneous morphology of the non-thermal
  emission traced by fluorescent Fe K$\alpha$ line has changed since
  2012, revealing three bright clumps around the Arches cluster, which
  are detected in 2015.

\item With long \xmm\ observations in 2015, we confirmed the declining
  trend of the non-thermal emission of the cloud reported by C14. 

\item By fitting a simple broken linear trend to the Arches cloud
  emission, we showed that the constant emission of the cloud started
  to decline in 2007. The fading rate is consistent both in power-law
  continuum and 6.4~keV line flux, which indicates that most of the
  non-thermal continuum was linked to the fluorescent Fe K$\alpha$
  6.4~keV line emission on a long time scale.  The estimated half-life
  time decay, $\tau_{1/2} = 8.05 \pm 1.28$~years, is similar to that
  observed in several other molecular clouds in the Galactic Centre,
  including the giant molecular cloud Sgr~B2, which adds more evidence
  that the same reflection mechanism is operating in both these clouds
  and that the main illuminating front(s) has already started to leave
  these clouds.

\item We have obtained the non-thermal X-ray spectrum of the Arches
  cloud by carefully separating the signal from thermal emission of the cluster
  core. The spectrum is well described by a power-law continuum with
  $\Gamma\sim2$ and a Gaussian line at 6.4~keV. A more physically
  motivated model based on Monte Carlo simulations fits the spectrum
  equally well; however, limited statistics does not allow for a constraint
  on the reflection geometry (Appendix~\ref{app:chur}).

\item The equivalent width of the fluorescent Fe K$\alpha$ line
  $EW_{\rm 6.4\ keV}$ revealed a sharp decrease to $0.6-0.7$~keV in
  2015, compared to $EW_{\rm 6.4\ keV}=0.9\pm0.1$~keV observed over 13
  years since 2002. This could indicate a change in reflection to
  different line of sight or metallicity of the clumps or that the CR
  component has become more dominant.

\item The measured $2-10$~keV unabsorbed total flux of the Arches
  cluster, $1.50_{-0.25}^{+0.30}\times10^{-12}$\,\ergscm, is fully
  consistent with its normal state flux level observed by
  \cite{capelli11a}, and no outburst from the cluster was detected
  during our observation in 2015.

\item Using \xmm\ and \nu\ observations in 2015, we demonstrate
  (see Apendix~\ref{app:SgrA}) a strong correlation between the 6.4~keV
  line flux and the hard $10-20$~keV X-ray continuum emission of the
  Br1 and Br2 clouds of the `Bridge' region \citep{ponti2010}. The
  emission of MC2 cloud is dim, which confirms the decreasing trend
  reported in earlier works \citep{capelli2012,clavel2013}.
\end{itemize}

\section*{Acknowledgments}

This work has made use of data from the \nu\ mission, a project led by
the California Institute of Technology, managed by the Jet Propulsion
Laboratory and funded by the National Aeronautics and Space
Administration, and reobservations obtained with XMM-Newton, an ESA
science mission with instruments and contributions directly funded by
ESA Member States and NASA. The research has made use of the \nu\ Data
Analysis Software ({\sl nustardas}) jointly developed by the ASI
Science Data Center (ASDC, Italy) and the California Institute of
Technology (USA). GP acknowledges support by the Bundesministerium
f\"ur Wirtschaft und Technologie/Deutsches Zentrum f\"ur Luft- und
Raumfahrt (BMWI/DLR, FKZ~50~OR 1408 and FKZ~50~OR~1604) and the Max
Planck Society. RK acknowledges support from the Russian Basic Research
Foundation (grant 16-02-00294), the Academy of Finland (grant 300005)
and hospitality of the Tuorla Observatory, and thanks Eugene
Churazov for the Monte Carlo X-ray reflection model used in the paper.


\appendix

\section{Serendepitous detection of the Sgr~A complex}
\label{app:SgrA}

\begin{figure}
\includegraphics[width=\columnwidth]{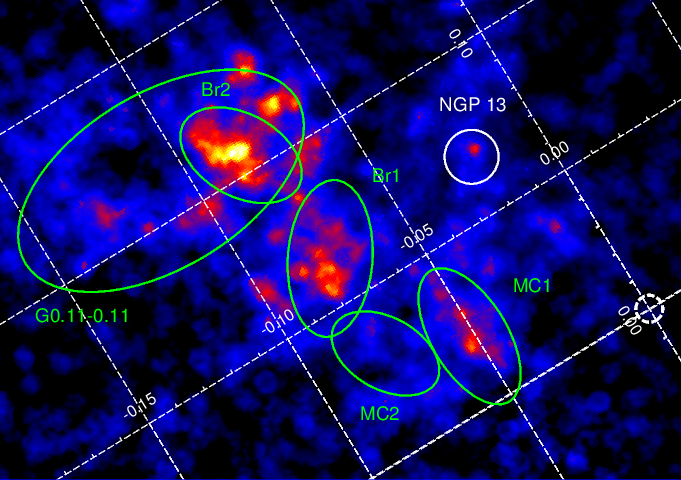}
\includegraphics[width=\columnwidth]{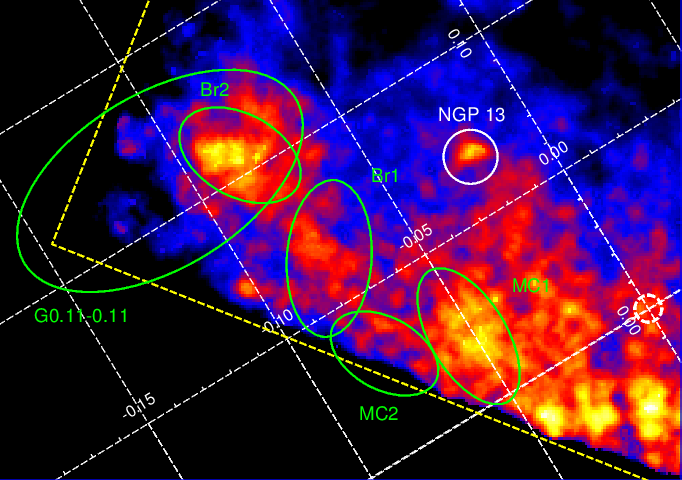}
\caption{{\it Upper panel:} \xmm\ image of the molecular zone complex
  region at 6.4~keV. {\it Lower panel:} \nu\ image of the same sky
  region in $10-20$~keV energy band. Green regions demonstrate
    the spatial extent of known illuminated clouds \citep{clavel2013}
    and white circle shows the position of hard X-ray point source
    NGP~13 \citep{hong2016}. The center of the Galactic coordinates is
    marked by the dashed circle. Yellow dashed lines delimit edges of
  the \nu\ image.}\label{fig:mc_clouds}
\end{figure}

Fig.~\ref{fig:mc_clouds} shows a part of CMZ region covered by \xmm\
and \nu\ observations of the Arches cluster, which allows us to
capture the present state of the cloud's surface brightness
distribution in 2015. The cloud designation is taken from
\cite{clavel2013} who analysed the {\it Chandra} data of the Sgr~A
complex, in particular the `Bridge' region \citep{ponti2010},
splitting it into the two clouds Br1 and Br2, which are the brightest
features in our observations. Fig.~\ref{fig:mc_clouds} demonstrates
the presence of a strong correlation between the 6.4~keV line flux (\xmm)
and hard $10-20$~keV X-ray continuum emission (\nu) of Br1 and
Br2. Unfortunately the region around the MC1 cloud is strongly
contaminated by ghost-rays (Fig.~\ref{fig:nustar:arches}).  Only partly
covered by the \nu\ FOV, the emission of the MC2 cloud is dim, both in
6.4~keV line and hard X-ray continuum, which is in agreement with the
decreasing trend reported in earlier works
\citep{capelli2012,clavel2013}. The detailed study of the Sgr~A
complex based on the current \xmm\ and \nu\ observations will be
presented in a separate paper.

\section{X-ray spectrum extraction from a 2D image}
\label{app:2d}

In the following procedure we repeat a simillar two-dimensional (2D)
image analysis as done by \cite{krivonos2016}, who analysed \nu\ data
of the local Seyfert 2 active galactic nucleus (AGN) NGC~5643 and
successfully decoupled partially confused spectra of the AGN core and
an ultra-luminous source located in the same galaxy.

\begin{figure}
\includegraphics[width=\columnwidth]{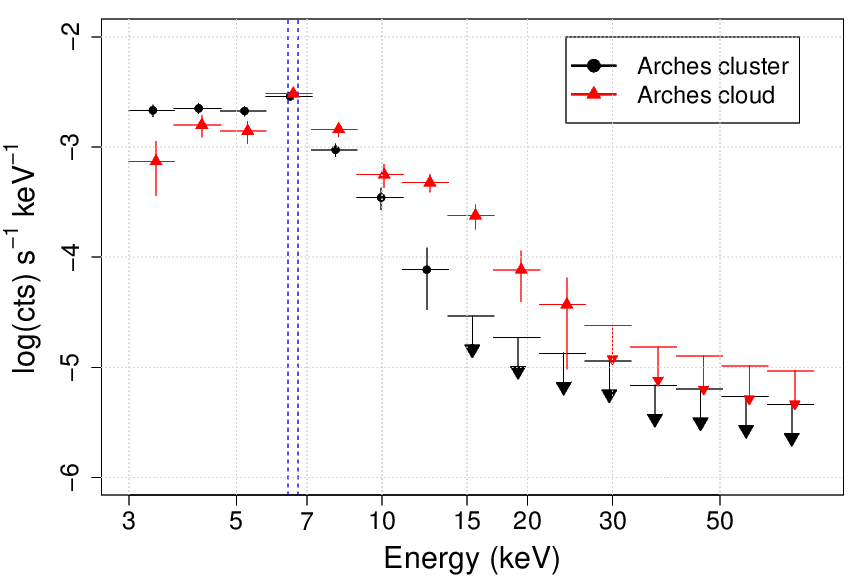}
\caption{Spatially resolved \nu\ spectra of the Arches cluster core
  (black) and cloud (red) X-ray emission. The position of Fe~K$\alpha$
  6.4 and 6.7~keV are marked by vertical dashed lines, as in the
  NuSTAR standard spectra these two lines cannot be separated by our
  technique.  The upper limits are 1$\sigma$
  errors.}\label{fig:nustar:2d}
\end{figure}

We first combined the FPMA data from both \nu\ observations into sky
mosaics in 15 energy bands logarithmically covering the \nu\ working
energy range $3-79$~keV. Each sky image was analysed with the {\sc sherpa}
package. The spatial model of the Arches cluster includes two 2D
Gaussians. The first represents the cluster's core emission, with the
position fixed at core centroid coordinates measured in K14 (see
Sect.~\ref{sec:data:nustar}) and width fixed at $4''$ FWHM (PSF
smearing effect). Similar to that, the second Gaussian was aligned
with the corresponding ``halo'' Gaussian component used in K14 to
describe the cloud emission, setting the position at
R.A.=17$^h$45$^m$50.62$^s$, Dec.=$-28^{\circ}49'47.17''$ and FWHM
model parameter at $72''.4$. Despite the fact that the morphology of
the cloud emission has been changed since 2012, the ``halo'' Gaussian
still covers the spatial extent of the molecular cloud region. The
amplitudes of the 2D Gaussians were free parameters. The background
term was estimated in the annulus $70''<R<130''$ shown in
Fig.~\ref{fig:nustar:arches}. By running the fitting procedure in each
of the 15 energy bands, we extracted the detector Pulse Height
Amplitude (PHA) spectrum of both spatial components. The corresponding
Redistribution Matrix File (RMF), which maps from energy space into
PHA space, was simply adopted from standard spectral analysis with
{\it nuproducts} (Sect.~\ref{sec:data:nustar}) and rebinned with {\it
  rbnrmf} tool of {\sc HEASOFT}~6.19 package. Following the approach
described in \cite{krivonos2016} we calibrated Auxiliary Response File
(ARF) utilizing the \nu\ data of a bright source with a known
spectrum. To this end we used a 20~ks observation (ObsID: 80001003002)
of the MeV Blazar PMN~J0641-0320 with a very hard power-law spectrum of
$\Gamma\approx1$ detectable up to $\sim80$ keV \citep{marco2016}. Here
we assume that the ARF calibrated for a point-like source is suitable for
the extended emission of the Arches cloud. Given the limited
statistics of the \nu\ Arches cluster observations, the deviations are
within the uncertainties.

Spatially decoupled spectra of the Arches cluster core and extended
cloud emission are shown in Fig.~\ref{fig:nustar:2d}. X-ray emission
of the cluster contains an excess in the $5.8-7.2$~keV range,
compatible with $\sim6.7$~keV line and rapidly drops above
$\sim10$~keV as expected for thermal emission with
$kT\approx2$~keV. Non-thermal emission of the extended cloud component
apparently includes excess around $6.4$~keV and dominates above
$10$~keV. This is consistent with what was inferred from the \nu\
images in $3-10$, $6-7$, and $10-20$~keV (K14).

\section{X-ray reflection model}
\label{app:chur}

The X-ray reflection model, based on Monte Carlo simulations
of the radiative transfer in the gas cloud, was initially developed
for calculation of an X-ray albedo of the Earth atmosphere
\citep{churazov2008}. The model was later modified to the geometry
more appropriate for illumination of GC molecular clouds. Namely, a
uniform spherical cloud is illuminated by an external source with a
given spectrum \citep{churazov2016}. The spectrum emerging from the
molecular cloud depends on several parameters: the slope ($\Gamma$) of
the incident power-law spectrum, the cloud optical depth to Thomson
scattering $\taut=\sigmat(2n_{{\rm H}_2})r$, where $n_{{\rm H}_2}$ is
the number density of hydrogen molecules and $r$ is the cloud radius,
the iron abundance relative to solar, the scattering angle ($\theta$)
for photons travelling from \sgra\ to the Arches cloud and then to the
observer, and the ISM line-of-sight column density toward the cloud
($N_{\rm H}$).

\begin{figure}
\includegraphics[width=\columnwidth]{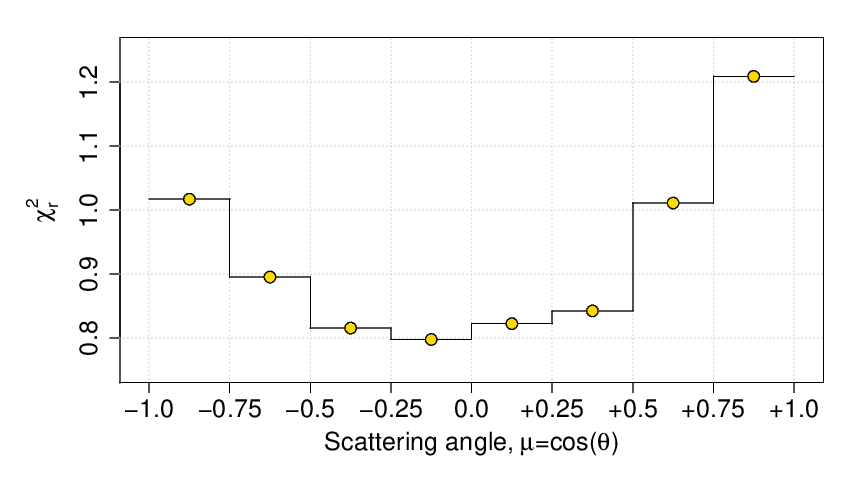}
\caption{Reduced $\chi^2$ fit statistics for 14 d.o.f. of the Arches
  cloud \nu\ spectrum fitted with the X-ray reflection model as a
  function of scattering angle $\theta$ expressed in
  $\mu=\cos(\theta)$.}\label{fig:nustar:2d:chur}
\end{figure}

\begin{figure}
\includegraphics[width=\columnwidth]{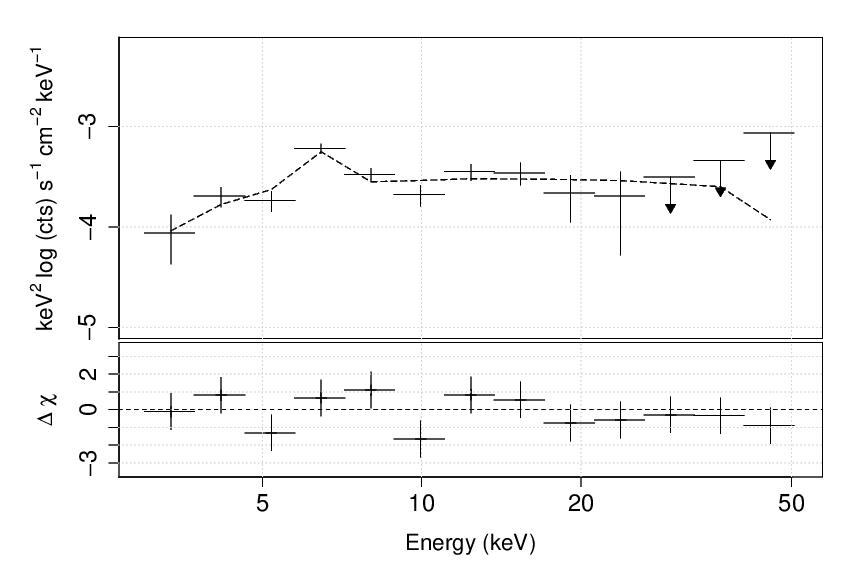}
\caption{Spatially resolved \nu\ spectrum of the Arches cloud
  modelled with X-ray reflection model.}\label{fig:nustar:2d:cloud:chur}
\end{figure}

In the current model setup we fixed the power-law slope of the incident
\sgra\ spectrum at $\Gamma=2$, consistent with a) what was measured
above, b) Sgr~B2 X-ray reflection scenarious
\citep{revnivtsev2004a,terrier2010,shuo2015,walls2016}, and c)
observed for \sgra\ flares
\citep{porquet2003,porquet2008,nowak2012,barriere2014}. The optical
depth of the cloud was not constrained by the fit ($\taut<0.06$), and
was fixed at the lowest value allowed by the model $\taut=0.02$, which
is in turn consistent with \cite{capelli11b} measurements
$\taut=0.01-0.03$ of the North and South bright Fe~K$\alpha$ knots of
the Arches cloud. The metallicity was fixed to $Z=1.7 Z\sun$, in
accordance with the whole analysis of this paper. To cover a wide
range of scattering angles $\theta$ from $180^{\circ}$ to $0^{\circ}$
(backscattering), we built a set of eight models for different
$\mu=cos(\theta)$, ranging from -1.0 to 1.0 with step of
0.25. Fig.~\ref{fig:nustar:2d:chur} shows fit statistics for the
models as a function of $\mu$. As seen from the figure, the quality of
the data formally allows wide range of the scattering angles $\theta$,
however revealing local extremum at $\mu=-0.25..0.0$
($\theta=90-105^{\circ}$), suggesting the cloud be located in or
slightly in front of the \sgra\ plane. The corresponding spectal fit
is shown in Fig.~\ref{fig:nustar:2d:cloud:chur}.

\bsp	
\label{lastpage}
\end{document}